\documentclass[useAMS,usenatbib]{mnras}
\usepackage[T1]{fontenc}
\pdfoutput=1
\usepackage{amsmath}
\usepackage{amssymb}
\usepackage{graphicx}
\usepackage{wasysym}
\usepackage{esint}
\usepackage[authoryear]{natbib}
\usepackage[utf8]{inputenc}
\usepackage{hyperref}
\usepackage{xspace}
\usepackage{newtxtext,newtxmath}
\usepackage{xcolor}
\usepackage[normalem]{ulem}

\newcommand{\arepo}{\textsc{Arepo}\xspace}
\newcommand{\Fermi}{\textit{Fermi}\xspace}
\defcitealias{2021WerhahnII}{Paper II}
\defcitealias{2021WerhahnIII}{Paper III}
\makeatletter

\pdfpageheight\paperheight
\pdfpagewidth\paperwidth

\title[CRs and non-thermal emission in galaxies I.]{Cosmic rays and non-thermal emission in simulated galaxies. I. Electron and proton spectra compared to Voyager-1 data}

\author[M. Werhahn et al.]{Maria Werhahn,$^{1,2}$\thanks{E-mail:
mwerhahn@aip.de} Christoph Pfrommer,$^{1}$ Philipp Girichidis,$^{1}$ Ewald Puchwein,$^{1}$ \newauthor R\"udiger Pakmor$^{3}$ 
\\
$^{1}$Leibniz-Institut f\"ur Astrophysik Potsdam (AIP), An der Sternwarte 16, 14482 Potsdam, Germany\\
$^2$Institut f\"ur Physik und Astronomie, Universit\"at Potsdam, Karl-Liebknecht-Str.\,24/25, 14476 Golm, Germany\\
$^{3}$Max-Planck-Institut f\"ur Astrophysik, Karl-Schwarzschild-Str. 1, 85741 Garching, Germany}
\usepackage{url}

\@ifundefined{showcaptionsetup}{}{%
 \PassOptionsToPackage{caption=false}{subfig}}
\usepackage{subfig}
\makeatother

\usepackage{babel}
\begin{document}
\date{Accepted 20XX . Received 20XX}

\maketitle
\pagerange{\pageref{firstpage}--\pageref{lastpage}} \pubyear{2020}

\label{firstpage}
\begin{abstract}
Current-day cosmic ray (CR) propagation studies use static Milky-Way models and fit parametrized source distributions to data. Instead, we use three-dimensional magneto-hydrodynamical (MHD) simulations of isolated galaxies with the moving-mesh code \arepo that self-consistently accounts for hydrodynamic effects of CR protons. In post-processing, we calculate their steady-state spectra, taking into account all relevant loss processes. We show that this steady-state assumption is well justified in the disc and generally for regions that emit non-thermal radio and gamma rays. Additionally, we model the spectra of primary electrons, accelerated by supernova remnants, and secondary electrons and positrons produced in hadronic CR proton interactions with the gas.  We find that proton spectra above 10~GeV only weakly depend on galactic radius, while they acquire a radial dependence at lower energies due to Coulomb interactions. Radiative losses steepen the spectra of primary CR electrons in the central galactic regions while diffusive losses dominate in the outskirts. Secondary electrons exhibit a steeper spectrum than primaries because they originate from the transported steeper CR proton spectra. Consistent with Voyager-1 and AMS-02 data, our models (i) show a turn-over of proton spectra below GeV energies due to Coulomb interactions so that electrons start to dominate the total particle spectra and (ii) match the shape of the positron fraction up to 10~GeV. We conclude that our steady-state CR modeling in MHD-CR galaxy simulations is sufficiently realistic to capture the dominant transport effects shaping their spectra, arguing for a full MHD treatment to accurately model CR transport in the future.
\end{abstract}

\section{Introduction}
Relativistic particles (so called CRs) are extremely rare in the interstellar medium (ISM): only one in $\sim 10^9$ particles is a CR particle. The CR population is mostly composed of protons with a small admixture of heavier nuclei, electrons and positrons. Despite the rarity of these highly energetic particles in terms of number density, their energy density is comparable to the thermal, magnetic and kinetic counterpart \citep{1990BoularesCox,2013Zweibel}. Hence, they inevitably play a crucial role in galaxy formation and evolution, e.g., by driving galactic winds and regulating star formation. This has been suggested in several different settings, such as one-dimensional (1D) flux tube models \citep{1991Breitschwerdt,1996Zarakashvili,1997Ptuskin,2008Everett,2010Samui,2016Recchia} or three-dimensional simulations of the ISM \citep{2013Hanasz, 2016Girichidis,2016Simpson,2018Farber}. In addition, CR-hydrodynamic simulations of forming galaxies in isolation \citep{2008Jubelgas,2012Uhlig,2013Booth,2014SalemBryan,2016bPakmor_winds,2017Ruszkowski, 2017bPfrommer,2018Jacob,2020DashyanDubois} or in cosmological environments \citep{2014SalemBryanHummels, 2020Buck,2020aHopkins,2020bHopkins} have shown the relevance of CRs in regulating the star formation rate (SFR) of galaxies and their ability to launch galactic winds. Up to PeV particle energies, CR protons are assumed to be mostly accelerated at the remnants of supernovae (SNe) and therefore of Galactic origin \citep[e.g.][]{1964ocr..book.....G,1985Jokipii}. The mechanism of diffusive shock acceleration leads to a distribution of CRs that can be expressed as a power law in momentum \citep{1987BlandfordEichler}. However, the locally observed spectrum is different from the freshly injected one, in terms of spectral distribution as well as composition, enabling theorists to infer CR propagation and interactions throughout the ISM. In particular, the abundance of secondary particle species that are created via CR interactions with the ISM provide insight into the propagation mechanisms \citep{2007Strong, GrenierBlackStrong2015}.

The strong connection between CRs and the physical properties of a galaxy can be deduced from observations of their non-thermal emission in radio \citep{1971VanDerKruit, 1992Condon, 2003Bell} and gamma rays  \citep{2012AckermannGamma, 2016RojasBravo,2017Linden} that tightly correlate with indicators of star formation. Since core-collapse SNe explode only $\sim5-30\,\mathrm{Myr}$ after star formation, a fresh population of CRs is closely connected to the existence of a young stellar population, whose ultraviolet (UV) radiation typically gets absorbed by dust and re-emitted in the far-infrared (FIR). Thus, under the assumption of calorimetry, i.e. that CRs lose most of their energy due to emission before they can escape, the FIR luminosity of star forming galaxies is expected to be strongly connected to the non-thermal emission arising from their CR population \citep{1994Pohl,1989Voelk,1996Lisenfeld}. Indeed, the radio synchrotron emission from CR electrons is found to be linearly correlated with the FIR luminosity of star forming galaxies. The correlation holds over many orders of magnitude from dwarf galaxies up to strongly starburst systems with a remarkably low scatter. This challenges our understanding of the processes in place, and needs fine-tuning of several parameters in order to explain the observations \citep{2010Lacki}.
In this context, the relevance of  primary versus secondary electrons has been discussed, the latter being claimed to play an important role especially in highly star-forming galaxies. 

Similarly, the gamma-ray luminosities are found to strongly correlate with the FIR luminosities of star forming galaxies, but deviate from this relation at low SFRs \citep{2020Ajello}. This has been attributed to non-radiative losses of CRs coming into play in low-density galaxies, where the calorimetric assumption can not be fulfilled \citep{2007Thompson,2010Strong,2011Lacki,2017bPfrommer,2020Kornecki}.

There are a number of numerical propagation models in the literature, e.g. GALPROP \citep{1998StrongMoskalenko}, DRAGON \citep{2008Evoli}, PICARD \citep{2014Kissmann} and Usine \citep{2020Maurin}.  They solve the CR propagation equation on a grid while parametrizing the distribution of supernova remnants (SNR) with various input parameters, while aiming to match all constraints given by the observations of CR nuclei, electrons and positrons, as well as the observed gamma-ray and synchrotron emission of our galaxy.
However, these models combine different models for the source distributions and independent inferences of the density and magnetic field of the Milky Way. As such, they are not self-consistently emerging from an MHD simulation, which limits their predictive power.

In an extra-galactic context, it is common to use one-zone leaky-box models of star-forming galaxies in order to explain various aspects of their observed non-thermal emission quantities \citep[e.g.\ ][]{2004Torres,2010Lacki,2011Lacki,2013Yoast-Hull}. These models prescribe one galaxy with a characteristic scale height, a fixed magnetic field and gas density, motivated by observations. Additionally, \cite{2016Heesen} and \cite{2019Miskolczi} use 1D CR transport models in order to explain the observed radio emission in the halos of star-forming spiral galaxies. However, there is no contact made to CR propagation models in the Milky Way, that probe the same underlying physics, and these one-zone models require fitting a number of parameters that are observationally not constrained.

We aim to develop a complementary approach to previous models by performing three-dimensional MHD simulations of forming galaxies. 
These simulations include CR protons as a relativistic fluid that is dynamically coupled to the MHD equations in the advection-diffusion approximation. We inject CRs at SNe that trace the sites of active star formation in the simulations. The dynamical impact of CR protons is thus included in the evolution of these galaxies. The spectral details of the CR protons as well as the CR electron and positron physics are investigated in post-processing. 

This paper is the first of a series of three papers. Here, we focus on studying the spectra of CR protons, primary and secondary electrons, that we assume to be in steady-state in each computational cell. Our approach is aimed to be predictive, since our source function is not tuned in order to exactly reproduce the observations, but results from modeling star-formation and CR physics self-consistently in our MHD simulations. Hence, we obtain observables that can be related to recent measurements of the Milky Way, such as the spectra of CR protons and electrons, as well as the fraction of positrons. The aspects of the resulting non-thermal emission will be studied in two following papers. \citeauthor{2021WerhahnII} (\citeyear{2021WerhahnII}, hereafter Paper II) provides insights into the gamma-ray emission from CRs in simulated galaxies, both in terms of total luminosities and spectral energy distributions and \citeauthor{2021WerhahnIII} (\citeyear{2021WerhahnIII}, hereafter Paper III) analyses the radio-synchrotron emission from the primary and secondary electron population.

We present our simulations in Section~\ref{sec: simulations}. The modeling of CRs is explained in Section~\ref{sec: CR-modeling}, where we detail how we obtain the spectra of CR protons, primary electrons and model the production of secondary electrons and positrons. Our resulting maps and spectra are discussed in Section~\ref{sec: results}, where we compare the latter to observations. Finally, we summarise our findings in Section~\ref{sec: summary and conclusion}. Appendix~\ref{appendix: loss rates and normalisation} provides more details on CR loss rates and the normalization of primary electron and proton spectra. Additionally, in Appendix~\ref{appendix: electron source fct. and parametrizations} we describe the production of secondary electrons and positrons, show our parametrization of the cross section of pion production, and compare the electron source function to an analytical approximation.

\section{Simulations} \label{sec: simulations}

\begin{table}
\caption{Overview of the different simulations.}
\label{tab:simulations-overview}
\begin{tabular}{lcccc}
\hline
 $M_{200}\,[\mathrm{M_{\odot}}]$ & $c_{200}$ & $\zeta_{\mathrm{SN}}$ & $B_0\,[\mathrm{G}]$& notes \\
\hline
$10^{12}$ & $7$   & $0.10$ & $10^{-10}$&\\
$10^{12}$ & $7$   & $0.05$ & $10^{-10}$& fiducial galaxy\\
$10^{12}$ & $12$ & $0.05$ & $10^{-10}$&\\
\hline
\end{tabular}
\end{table}

\begin{figure*}
\begin{centering}
\includegraphics[]
{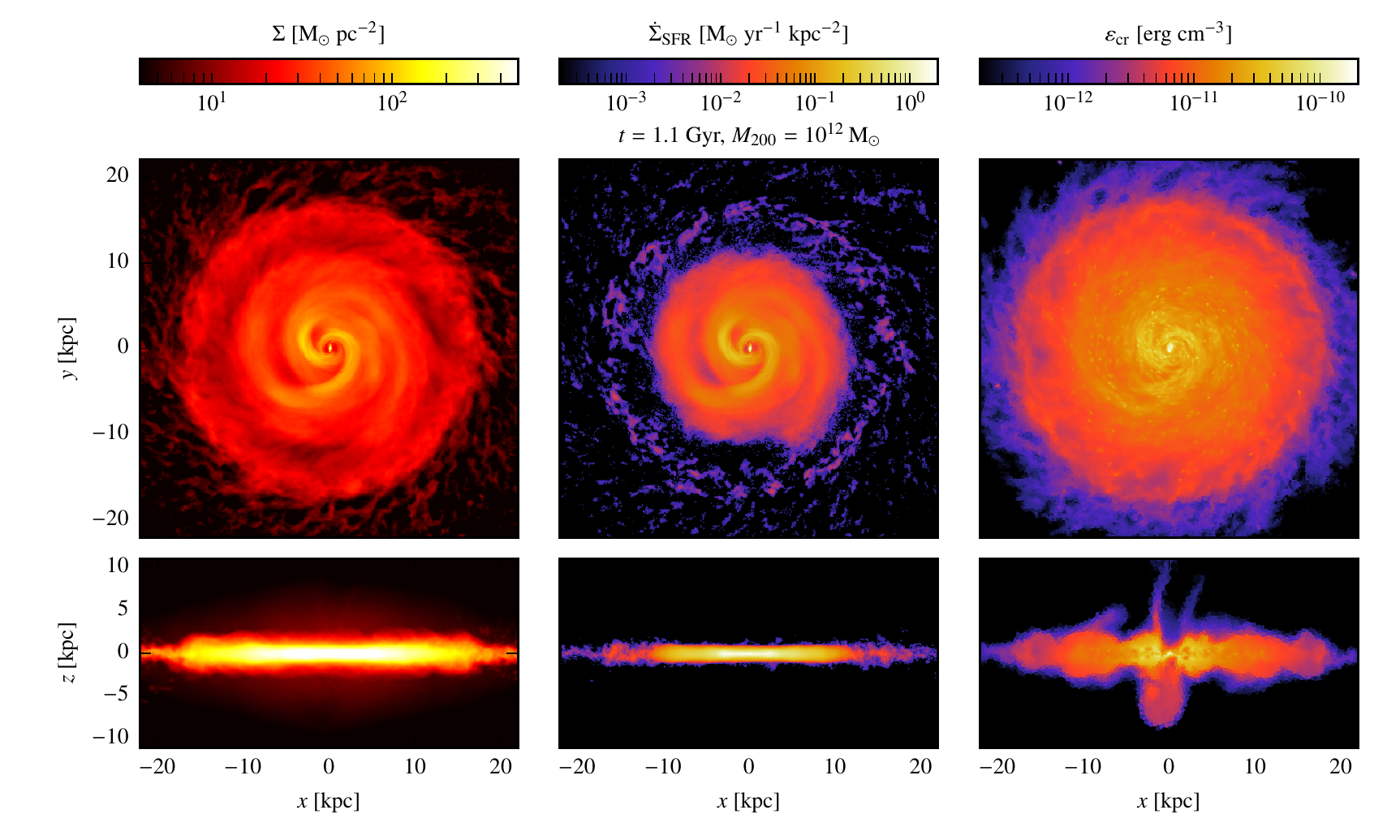}
\caption{Left to right, we show projected maps of the gas surface density $\Sigma$, SFR density $\dot{\Sigma}_{\mathrm{SFR}}$ and slices of the cosmic ray energy density $\varepsilon_{\mathrm{cr}}$ at 1.1 Gyr. Shown are face-on views (top panels) and edge-on views (bottom panels) for a galaxy with a halo mass of $M_{200}= 10^{12}\,\mathrm{M_{\odot}}$, concentration $c_{200}=7$, initial magnetic field $B_0=10^{-10}\,\mathrm{G}$ and CR acceleration efficiency $\zeta_{\mathrm{SN}}=0.05$. In the following, we refer to this configuration as our fiducial galaxy.}
\label{maps-properties}
\par\end{centering}
\end{figure*}

\begin{figure*}
\begin{centering}
\includegraphics[]{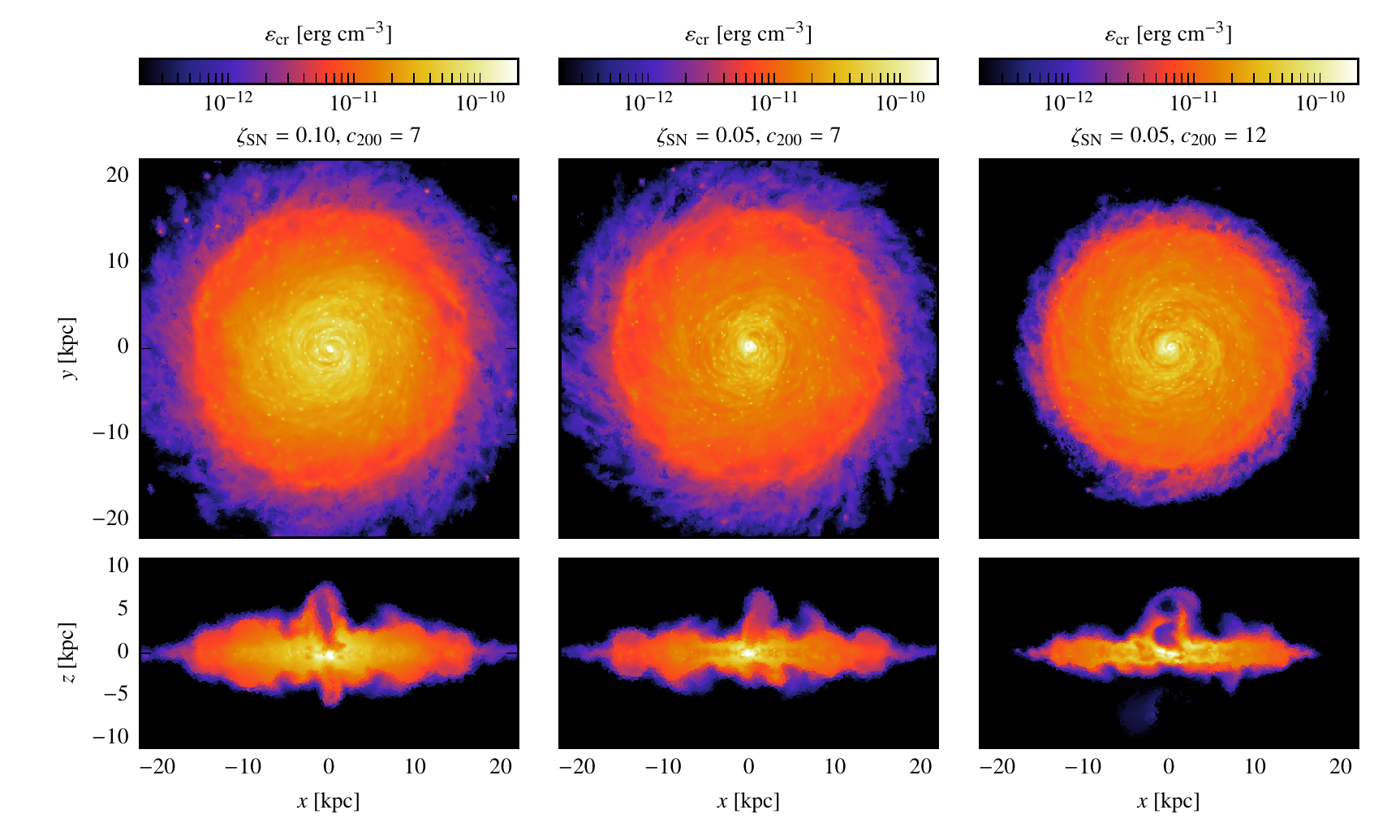}
\caption{We show face-on views (top panels) and edge-on views (bottom panels) of slices of the CR energy density $\varepsilon_\mathrm{cr}$ through three different simulations with the same halo mass $M_{200}= 10^{12}\,\mathrm{M_{\odot}}$ at 1\,Gyr, but different concentration parameters $c_{200}$ and injection efficiencies $\zeta_{\mathrm{SN}}$ as indicated in the panels. The central panel shows our fiducial galaxy that we will analyse in the following.}
\label{maps-properties-c7-c12}
\par\end{centering}
\end{figure*}

We perform MHD simulations of isolated galactic discs with the moving mesh code \arepo \citep{2010Springel,  2016aPakmor}, which simulates magnetic fields with ideal MHD \citep{2013Pakmor}. The simulations performed in this study are similar to those in \citet{2017bPfrommer} and use the one-moment CR hydrodynamics algorithm \citep{2016cPakmor,2017aPfrommer}. In order to cover the entire mass spectrum of galaxies from dwarfs to Milky Way-like galaxies, we simulate dark matter halo masses ranging from $M_{\mathrm{200}}=10^{10}$ to $10^{12}\, \mathrm{M_{\odot}}$. The gas cloud is initially assumed to be in hydrostatic equilibrium with the halo. It contains $10^7$ gas cells, each carrying a target mass\footnote{We enforce that the gas mass of all Voronoi cells remains within a factor of two of the target mass by explicitly refining and de-refining the mesh cells.} of $155\,\mathrm{M_{\odot}}\times M_{200}/(10^{10}\,\mathrm{M_{\odot}})$, embedded in a dark matter halo that follows an NFW profile \citep{1997Navarro}. This is characterised by a concentration $c_{200}=r_{200}/r_{\mathrm{s}}$, where $r_{\mathrm{s}}$ is the characteristic scale radius of the NFW profile and the radius $r_{200}$ encloses a mean density that corresponds to 200 times the critical cosmic density. We  assume a baryon mass fraction of $\Omega_{\mathrm{b}}/\Omega_{\mathrm{m}}=0.155$ and assign initial angular momentum to the halo, which is parametrized by a spin parameter $\lambda =j/(\sqrt{2}R_\rmn{vir}V_\rmn{vir})=0.05$, where $j=J/M$ is the specific angular momentum of the halo and $R_\rmn{vir}$ and $V_\rmn{vir}$ denote the virial radius and velocity of the halo, respectively. Our choice of  the radial distribution of $\lambda$ is in agreement with results from full cosmological simulations \citep{Bullok2001}. We switch on cooling at $t=0$ which is fastest in the center. This causes loss of pressure support and infall of the gas while it conserves its specific angular momentum. As a result, a galactic disc starts to form from the inside out. While this problem set-up is axisymmetric, the simulation result is not axisymmetric due to the probabilistic star formation model that we will explain in the following.

The simulations follow a simplified model of star formation and instantaneous core-collapse SN feedback \citep{2003SpringelHernquist}, in which regions above a critical threshold density are stochastically forming stars with an expectation value consistent with the observed \citet{1998Kennicutt} law. CRs are instantaneously injected at the SNe, and obtain a fraction $\zeta_{\mathrm{SN}}$ of the kinetic energy of the SN explosion. This implies that the CR energy gain of a cell with SFR $\dot{m}_\star$ is given by $\Delta E_{\mathrm{CR}}=\zeta_{\mathrm{SN}} \epsilon_{\mathrm{SN}} \dot{m}_\star \Delta t$, where $\epsilon_{\mathrm{SN}}=10^{49}\, \mathrm{erg}\, \mathrm{M_{\odot}^{-1}}$ is the released specific energy. 

We adopt an initial seed magnetic field before collapse of the gas cloud of  $B_0=10^{-10}\,\mathrm{G}$ and an injection efficiency of $\zeta_{\mathrm{SN}}=0.05$ and $0.10$. The injected CRs are advected with the gas, while adiabatic changes in the CR energy are taken into account. We also account for CR losses due to Coulomb interactions as well as hadronic losses as a consequence of inelastic collisions with the thermal ISM. Furthermore, we include anisotropic diffusion of CRs along the magnetic field, as described in \cite{2016cPakmor} and adopt a parallel CR diffusion coefficient along the local magnetic field of $D=10^{28}~\rmn{cm^2~s}^{-1}$.\footnote{This value is consistent with the recently discovered hardening of the logarithmic momentum slope of the CR proton spectrum at low Galactocentric radii, which is interpreted as a signature of anisotropic diffusion in the Galactic magnetic field \citep{Cerri2017,Evoli2017}. Using the flux of unstable secondary CR nuclei in recent AMS-02 data, which signals spallation processes in the ISM, the residence time of CRs inside the Galaxy can be constrained to yield identical values for the diffusion coefficient \citep{Evoli2019,2020Evoli}.} For halo masses of $10^{12}\, \mathrm{M_{\odot}}$, we adopt different values of the concentration parameter $c_{200}$ for the dark matter halo. 

Table~\ref{tab:simulations-overview} provides an overview of the different combinations of injection efficiency and concentration parameters that we use in our simulations of the $10^{12}\,\mathrm{M_{\odot}}$ halo, which we analyse here.
We focus on this Milky Way-sized galaxy in this paper, whereas the smaller galaxies will be analysed in \citetalias{2021WerhahnII} and \citetalias{2021WerhahnIII}. Figure~\ref{maps-properties} depicts a snapshot of the simulation with $10^{12}\,\mathrm{M}_\odot$ and $c_{200}=7$ after $t=1.1\,\mathrm{Gyr}$ of evolution. The morphologies of the gas column density $\Sigma$ and the SFR column density $\dot{\Sigma}_\mathrm{SFR}$, which are manifested by spiral structures, self-consistently determine the CR source distribution via our star formation model and hence give rise to a similar structure of the CR energy density. The CR pressure gradient was successful in driving an asymmetric outflow from the center that shows the largest SFR, which has evacuated underdense channels above and below the disc (visible in the  column density map).

\begin{figure*}
\begin{centering}
\includegraphics[]{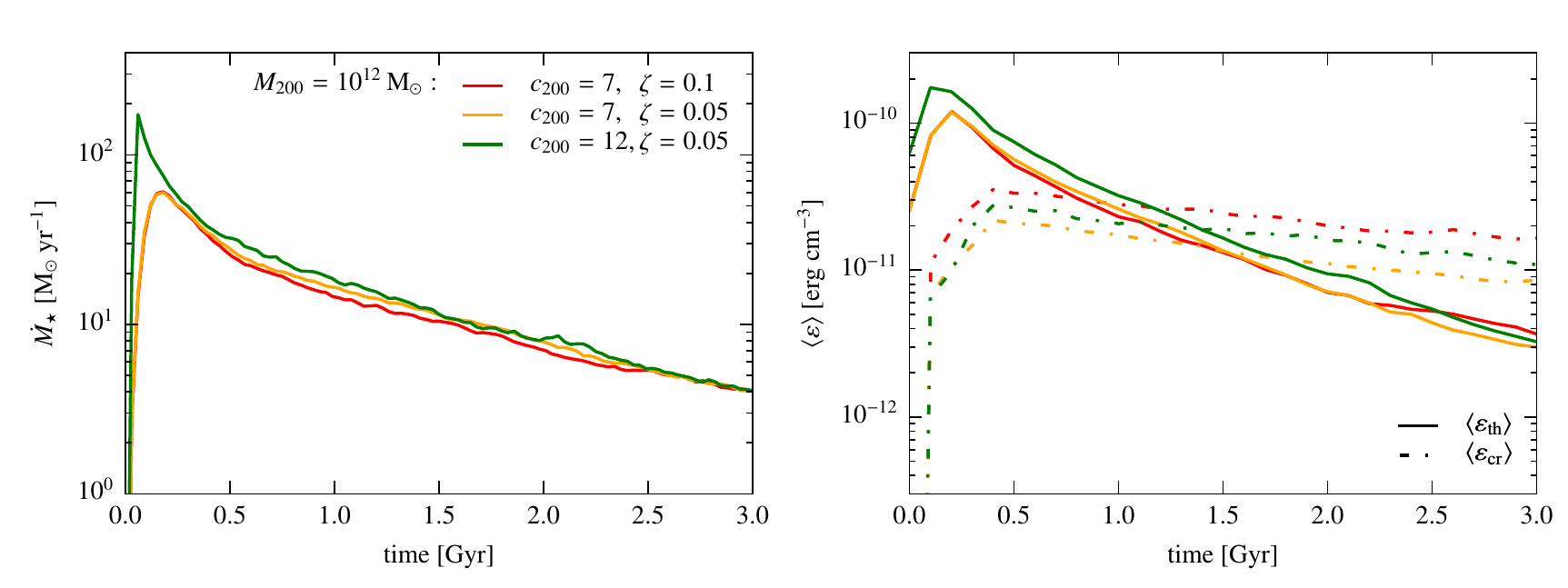}
\caption{Time evolution of the SFR (left-hand panel) and the volume-averaged thermal and CR energy densities in a disk of radius 10~kpc and total height 1~kpc (right-hand panel) for our Milky Way-like galaxy with a halo mass of $10^{12}~\mathrm{M}_{\odot}$. We vary halo concentration parameter $c_{200}=\{7,12\}$ and CR energy acceleration efficiency $\zeta=\{0.05,0.1\}$. Note that  our fiducial galaxy is characterised by the combination $c_{200}=7$ and $\zeta=0.05$ (shown in yellow).}
\label{time-evolution}
\par\end{centering}
\end{figure*}

Despite the axisymmetric problem setup, we have seen that the emergent galactic winds are not symmetric with respect to the disc. In order to understand the reason for this asymmetric outflow morphology, we study the dependence of the CR distribution on various parameters in Fig.~\ref{maps-properties-c7-c12}, which shows slices of the CR energy density. Prime among those parameter is i) the CR acceleration efficiency $\zeta_{\mathrm{SN}}$ that determines the amount of injected CR energy at SNe and hence sets the CR gradient strength that is able to drive galactic winds and ii) the halo concentration $c_{200}$ that determines the potential depth of the halo. Increasing values of $c_{200}$ imply a larger density and hence a deeper dark matter potential. Indeed, Fig.~\ref{maps-properties-c7-c12} shows that larger values of $c_{200}$ imply more compact discs and weaker outflows while larger values of $\zeta_{\mathrm{SN}}$ increase the outflow strengths. Hence, small changes in these parameters have large consequences on whether CRs can drive outflows in Milky Way-mass galaxies and shape the particular outflow morphologies. This is only possible because CR-driven winds are getting weaker towards the mass scale of Milky Way-mass galaxies \citep{2012Uhlig, 2018Jacob} so that the onset of a CR-driven wind represents an unstable phenomenon and critically depends on the acceleration strength (i.e., the CR gradient) and the gravitational attraction of the density of dark matter and the amount of stars, which increases with time and furthermore deepens the central potential. Hence, the asymmetry arises as an emergent phenomenon that is the result of the outflow taking the path of least resistance away from the galaxy, which may be blocked or obscured in one hemisphere of the galaxy.

Note that despite the different galactic outflow appearances and strengths, the global properties such as formed stellar mass or average thermal or CR energy are rather similar among these Milky Way-like models studied here. This is exemplified in Fig.~\ref{time-evolution} by analysing the SFRs (left-hand side) and volume-averaged thermal and CR energy densities in a disk of fixed radius and height for our three galaxies (right-hand side). The peak SFR increases by a factor of three when the halo concentration is increased from $c_{200}=7$ to $12$ due to the different compression upon the initial collapse, but this leaves little impact on the average thermal and CR energies. In particular, the latter quantity differs by approximately a factor of two, which exactly resembles the difference in energy injection efficiency at SNRs.

\begin{figure*}
\begin{centering}
\includegraphics[]
{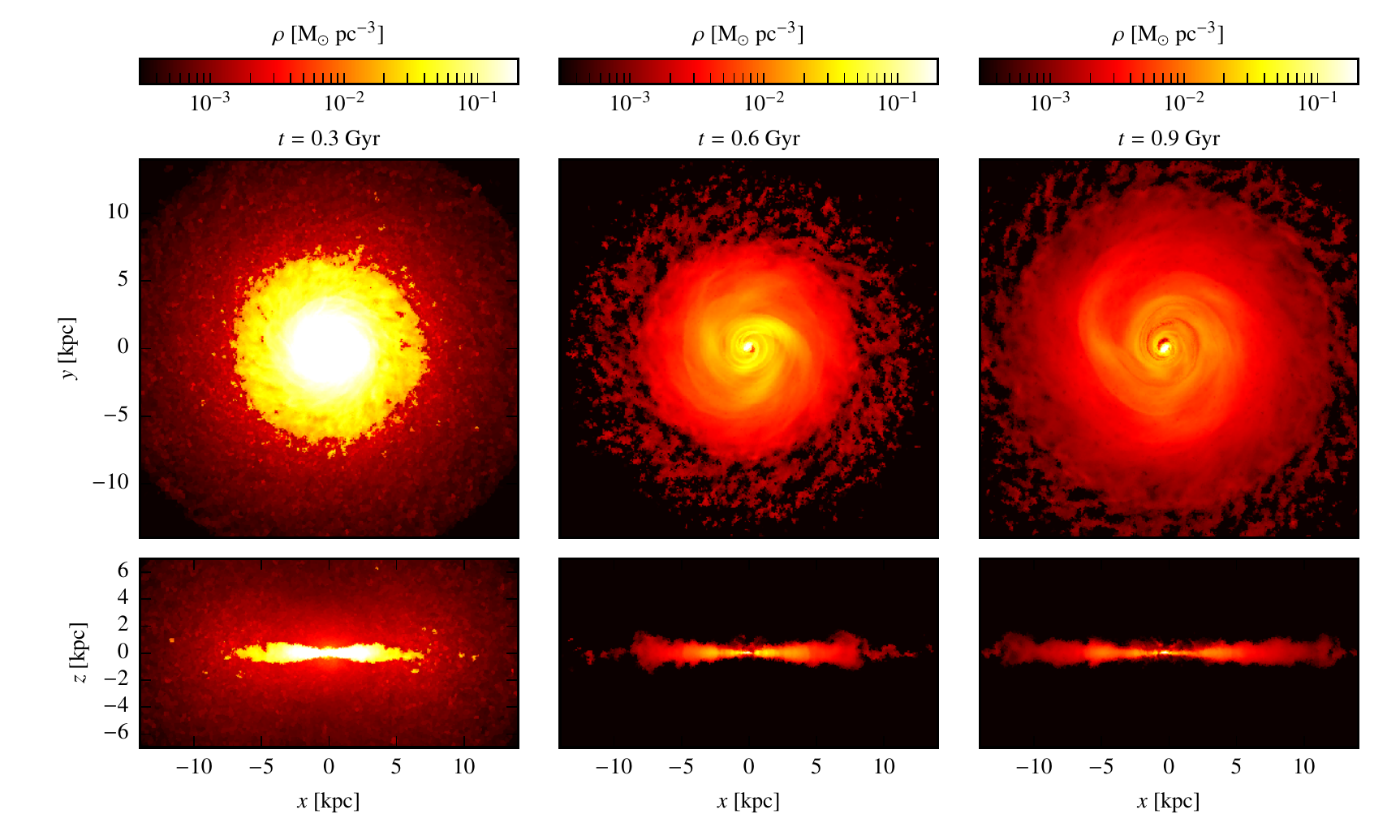}
\par\end{centering}
\begin{centering}
\includegraphics[]{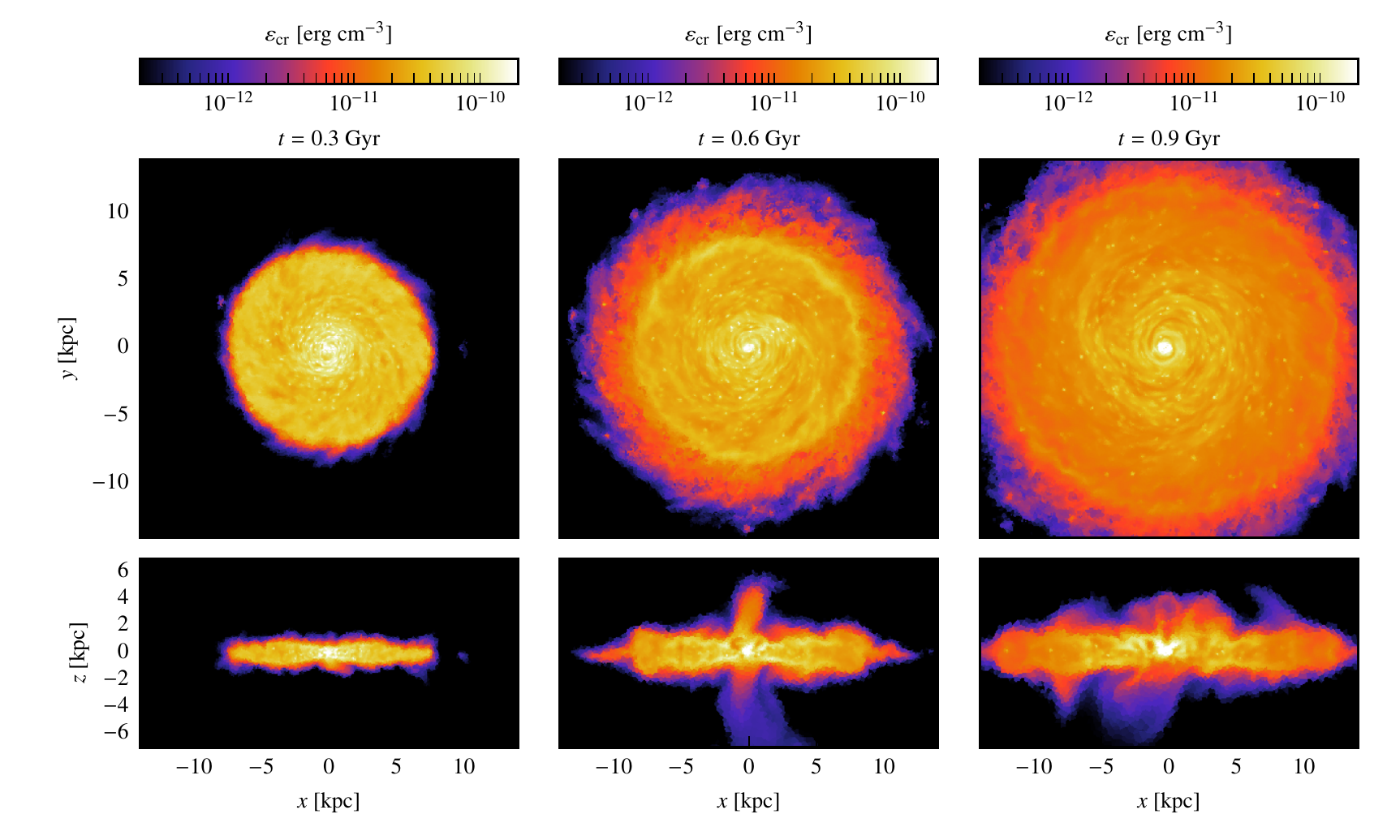}
\par\end{centering}
\caption{Temporal evolution of slices of the gas density $\rho$ (top panels) and slices of the cosmic ray energy density $\varepsilon_{\mathrm{cr}}$ (bottom panels) for a galaxy with a halo mass of $M_{200}= 10^{12}\,\mathrm{M_{\odot}}$, concentration $c_{200}=12$, initial magnetic field $B_0=10^{-10}\,\mathrm{G}$ and CR acceleration efficiency at SNe $\zeta_{\mathrm{SN}} =0.05$. Each panel shows the face-on view on top of the edge-on view, respectively.}
\label{maps-time-evolution}
\end{figure*}

\begin{figure}
\begin{centering}
\includegraphics[]
{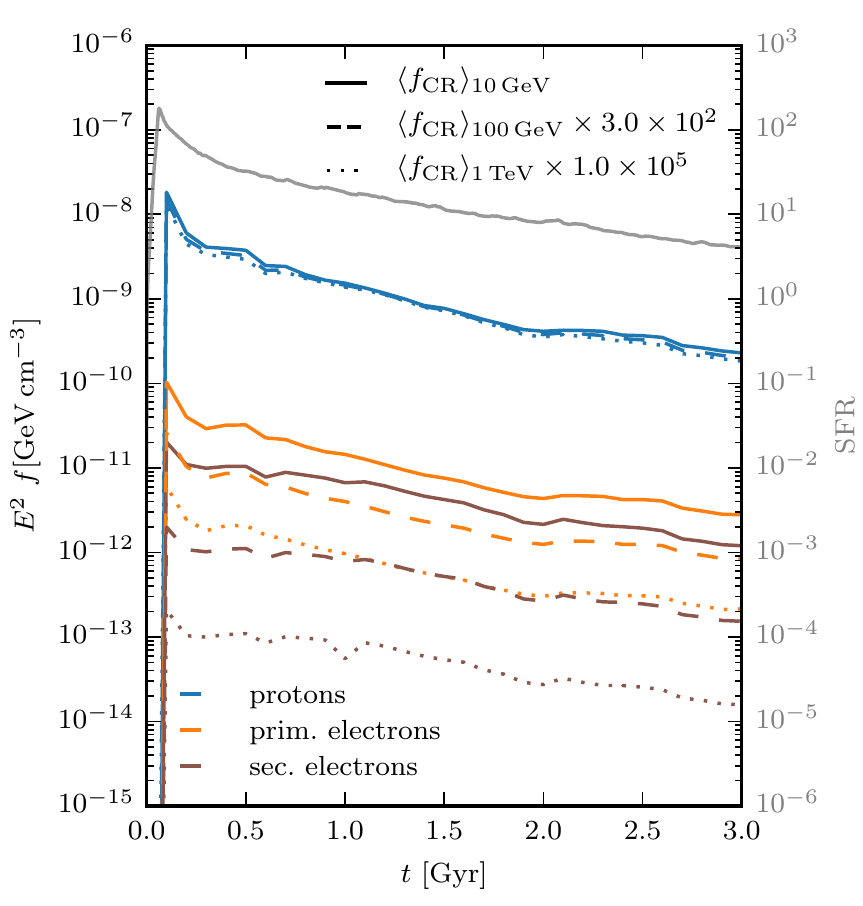}
\end{centering}
\caption{Evolution of the mean spectral energy density $f$ of each CR species at 10 GeV, 100 GeV and 1 TeV over time for the simulation shown in Fig.\ \ref{maps-time-evolution}. The spectra are averaged over the gas scale height and the radius which includes 99 per cent of the hadronic gamma-ray luminosity \citepalias[see][for the calculation of the non-thermal emission processes]{2021WerhahnII}. In addition, we show the SFR (grey line) as a function of time.}
\label{evolution_CR_energy_density}
\end{figure}

The time evolution of our simulations is exemplified in Figs.\,\ref{maps-time-evolution} and \ref{evolution_CR_energy_density} for a simulation with $M_{200}= 10^{12}\,\mathrm{M_{\odot}}$, $c_{200}=12$,  $B_0=10^{-10}$ and $\zeta_{\mathrm{SN}} =0.05$. In particular, maps of the gas density in Fig.~\ref{maps-time-evolution} show the formation of a rotationally supported disc at a few hundred Myrs after the initial gas cloud has started to collapse. Because CRs are injected at remnants of SN explosions, they reside in the star-forming disc before they are transported through advection or diffusion. For the parameters chosen here, the CR pressure in the disc is sufficiently high after 600 Myrs in order to bend and open up the toroidal magnetic field, enabling them to diffuse into the halo and drive an outflow. Due to the decreasing SFR and hence decreasing CR injection rate, the CR gradient weakens over time and the outflow eventually dissolves another 300 Myrs later.

The initial collapse of the gas cloud results in a short starburst, followed by an exponentially declining SFR, see Figs.~\ref{time-evolution}  and \ref{evolution_CR_energy_density}. Additionally in Fig.~\ref{evolution_CR_energy_density}, we show the mean energy density of CR protons, primary and secondary electrons at different energies (10 GeV, 100 GeV and 1 TeV respectively) as a function of time. They do not show any significant differences in the temporal evolution at the considered energies, but simply follow the evolution of the SFR. 
Here, we average over the radius, which includes 99 per cent of the hadronic gamma-ray emission, and the gas scale height $h_\rho$, where the gas density has dropped by an e-folding. The latter increases from $h_{\rho}=0.13\,$kpc to $h_{\rho}=0.83\,$kpc from $t=0.1$ to $t=2.3$~Gyr, before it approaches $h_{\rho}=0.73\,$kpc at $t=3$~Gyr. However, we note that our results do not depend on the specific choice of the averaging volume.

\section{Cosmic ray modeling} \label{sec: CR-modeling}
For each snapshot of our simulations, we model the CR spectra in terms of a cell-based steady-state approximation, which assumes that the considered loss processes occur on a timescale shorter than the timescale of the total change in simulated CR energy density in each cell so that CR sources and losses balance each other.

\subsection{Steady-State spectra}
We separately solve three diffusion-loss equations \citep[see e.g.][]{1964ocr..book.....G, 2004Torres} to obtain the equilibrium spectra for the spectral densities $f(E)=\mathrm{d}N/(\mathrm{d}E\,\mathrm{d}V)$, i.e. the number of particles per 
unit volume and unit energy, of CR protons, primary and secondary electrons, where $E$ denotes the total particle energy. It assumes that the injection of CRs, given by the source term $q(E)=\mathrm{d}N/(\mathrm{d}E\,\mathrm{d}V\,\mathrm{d}t)$, the CR production rate per unit volume and unit energy, is balanced by cooling, i.e., energy losses $b(E)=-\mathrm{d}E/\mathrm{d}t$, and escape from the system. The latter includes advective and diffusive losses, which are combined in an energy-dependent escape timescale $\tau_{\mathrm{esc}}=1/(\tau_{\mathrm{adv}}^{-1} + \tau_{\mathrm{diff}}^{-1})$. For each CR population, we solve the following equation
\begin{equation}
\frac{\mathrm{}f(E)}{\tau_{\mathrm{esc}}}-\frac{\mathrm{d}}{\mathrm{d}E}\left[f(E)b(E)\right]=q(E),
\label{eq:diff-loss-equ}
\end{equation}
which can be solved using the Green's function
\begin{equation}
G(E,E')=\frac{1}{b(E)}\exp\left(-\intop_{E}^{E'}\mathrm{d}y\frac{1}{\tau_{\mathrm{esc}}(y)b(y)}\right).\label{eq:green function diff. loss equation}
\end{equation}
We obtain the steady-state distribution $f(E)$ in each cell from an injected source function of CRs, $q(E')$, by integrating over the initial energy $E'$, i.e., 
\begin{equation}
f(E)=\intop_{E}^{\infty}\mathrm{d}E'q(E')G(E,E'),
\label{eq:N(E) from diff. loss equation}
\end{equation}
where $q(E) = q[p(E)] \rmn{d}p/\rmn{d}E$ and the injection spectra of CR electrons and protons are given in terms of a power-law in momentum,
\begin{align}
q_{i}(p_{i})\mathrm{d}p_{i} = C_{i} p_{i}^{-\alpha_{\mathrm{inj}}} \exp[-(p_i/p_{\mathrm{cut},i})^{n}]\mathrm{d}p_{i},
\label{eq: source fct. Q(p)}
\end{align}
where the normalised particle momentum is given by
\begin{align}
p_{i}=\frac{P_i}{m_{i}c}=\sqrt{\left(\frac{E_i}{m_i c^2}\right)^2 - 1},
\label{eq: momentum}
\end{align}
the subscript $i=\mathrm{e,p}$ denotes the CR species and $n=1$ for protons and $n=2$ for electrons \citep{2007Zirakashvili,2010Blasi}. The normalization $C_i$ is given in units of $\mathrm{s^{-1}\,cm^{-3}}$. We assume that both protons and primary electrons share the same injection spectral index of $\alpha_{\mathrm{inj}}=2.2$ \citep{2013LackiThompson} and assume cutoff momenta for protons, $p_{\mathrm{cut,p}}=1\,\mathrm{PeV}/m_{\mathrm{p}}c^2$  \citep{1990Gaisser}, and electrons, $p_{\mathrm{cut,e}}=20\,\mathrm{TeV}/m_{\mathrm{e}}c^2$ \citep{2012Vink}. While we inject CR protons at SNRs and transport their energy density with our MHD code (\citealt{2017aPfrommer}, see also Section~\ref{sec: simulations}) the source term for the electrons is unspecified in our formalism and may include all relevant sources (SNRs, pulsar-wind nebulae, gamma-ray binaries). In principle, we could implement different source spectra for leptonic and hadronic CRs but leave the investigation of this degree of freedom to future work. In order to solve for the spectral energy distribution, in practice we discretise the spectrum in logarithmically equidistant momentum bins and integrate Eq.~(\ref{eq:green function diff. loss equation}) for every momentum bin, before we perform the integration over the Green's function in Eq.~(\ref{eq:N(E) from diff. loss equation}) by using the trapezoidal rule.

\subsubsection{Energy losses and timescales}
The energy losses are evaluated locally in every cell, using the present physical properties of the cell. In order to separately solve Eq.~(\ref{eq:diff-loss-equ}) for CR protons and electrons, we have to consider the corresponding loss processes $b(E)=\dot{E}$ for each species.

The simulations already account for the CR proton losses. But in order to obtain a representation of the spectral distribution of CRs, we assume a steady-state and solve the diffusion-loss equation for CR protons. This might give us a modified spectral index in comparison to the injection index $\alpha_{\mathrm{p}}$ if an energy-dependent loss process dominates the cooling. Finally, we re-normalise the steady-state distribution to the simulated CR energy density in each computational Voronoi cell. We consider the following energy-loss processes of CR protons.

\begin{figure*}
\begin{centering}
\includegraphics[width=0.52\textwidth]{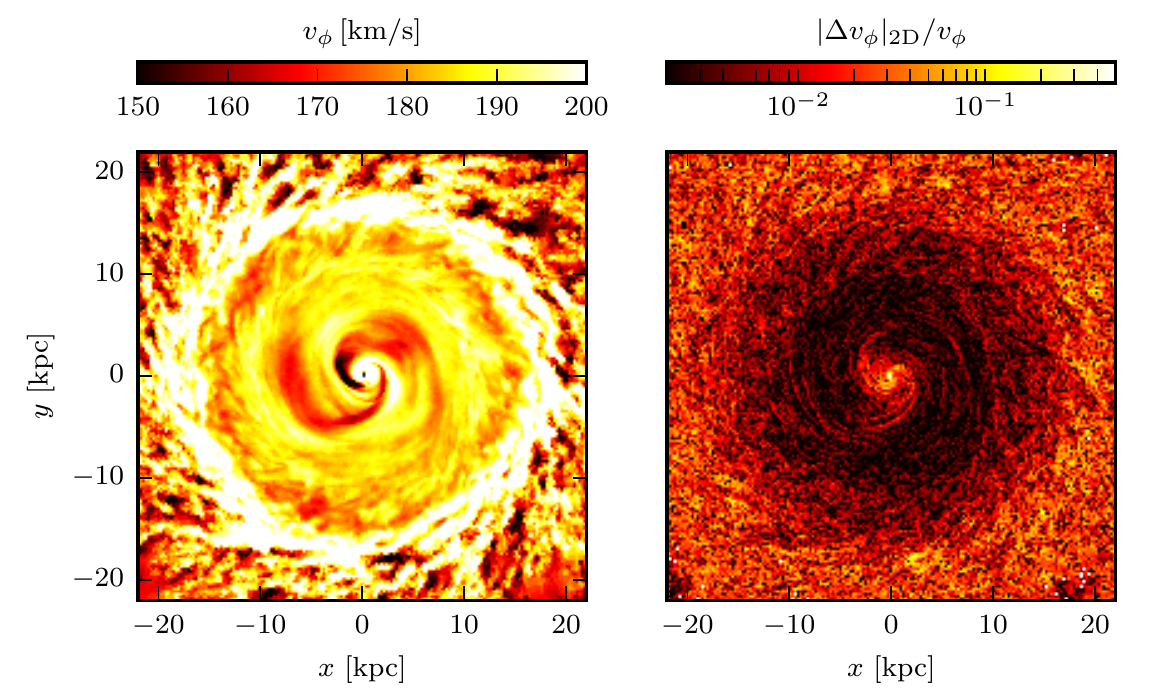}\includegraphics[width=0.42\textwidth]{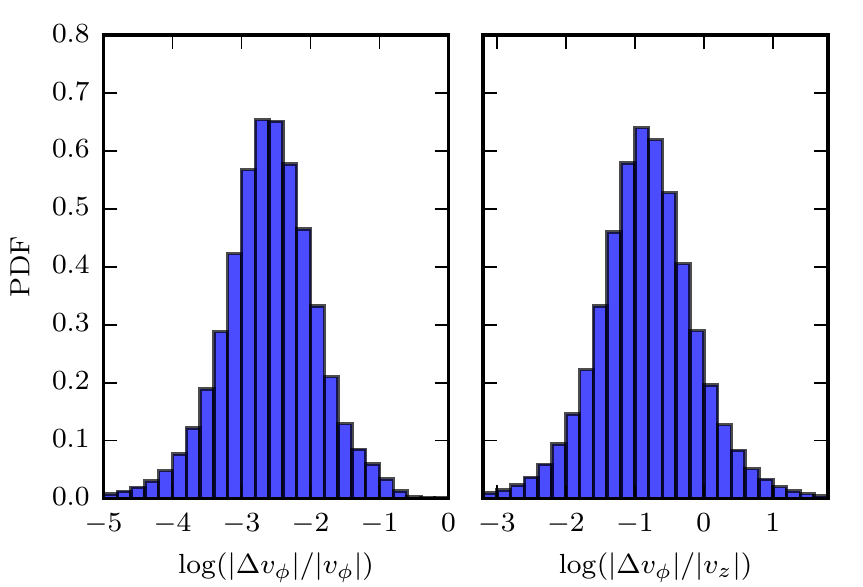}
\par\end{centering}
\caption{Face-on map of the azimuthal velocities $\varv_{\phi}$ within the disk of our fiducial galaxy at $t=1.1\,\mathrm{Gyr}$ (left-hand panel). The second panel shows the corresponding gradient map, where we multiply the absolute value of the gradient with the resolution of the 2D-map in order to obtain $\lvert \Delta \varv_{\phi}\rvert_\mathrm{2D}=\lvert\nabla \varv_\phi\rvert \Delta x$. In addition, the two right-hand panels show histograms of the azimuthal velocity difference of each cell relative to the mean velocity of its neighbouring cells in three dimensions, $|\Delta \varv_{\phi}|=|\varv_{\phi}-\langle \varv_{\phi}\rangle_{\mathrm{neighbours}}|$ relative to the absolute values of the azimuthal velocity ($|\varv_{\phi}|$, third panel) and vertical velocity, respectively ($|\varv_z|$, forth panel); see text for details.}
\label{fig:azimuthal-velocities}
\end{figure*}

\paragraph*{Protons.}
CR protons lose energy due to (i) hadronic interactions with the ambient medium, which produce pions (Eq.~\ref{eq: b_pi}) and (ii) through Coulomb interactions (Eq.~\ref{eq:b_Coulomb_p}), which heat the ISM. In order to avoid double-counting of adiabatic loss and gain processes we only account for this effect in our MHD simulations and neglect the spectral changes associated with this process. Modeling adiabatic spectral changes will require to follow the evolution of the spectral CR distribution in space and time in a galaxy simulation, which we postpone to future work \citep{2020MNRAS.491..993G}. Furthermore, we have to specify the characteristic timescale of losses due to escape. The residence time of CRs is determined by diffusion and advection, i.e., 
\begin{equation}
\tau_{\mathrm{esc}}^{-1}=\tau_{\mathrm{diff}}^{-1}+\tau_{\mathrm{adv}}^{-1}.\label{eq: timescale lifetime}
\end{equation}
We estimate the diffusion timescale by
\begin{align}
\tau_{\mathrm{diff}}=\frac{L_{\mathrm{CR}}^{2}}{D(E)}\propto E^{-0.5},
\end{align}
with the energy-dependent diffusion coefficient $D(E) = D_0 (E/E_0)^{0.5}$ as inferred from observed beryllium isotope ratios \citep{2020Evoli}, where $D_0=10^{28}\,\mathrm{cm^2~s}^{-1}$, $E_0 = 3$~GeV, and we adopt the diffusion length in each cell, $L_{\mathrm{CR}}=\varepsilon_{\mathrm{CR}}/\left|\nabla\varepsilon_{\mathrm{CR}}\right|$. The characteristic timescale of advection is calculated as 
\begin{align}
\tau_{\mathrm{adv}}=\frac{L_{\mathrm{CR}}}{\varv_{z}}.
\label{eq: t_adv}
\end{align}
Assuming that in our cell-based steady-state approximation the azimuthal fluxes in and out of the cell compensate each other, we only take the velocity $\varv_{z}$ of the cell in $z$-direction perpendicular to the disc into account for the estimation of the advection timescale. 
To justify this assumption, we show in Fig.~\ref{fig:azimuthal-velocities} a map of the azimuthal velocity field $\varv_{\phi}$ in the disc that only varies smoothly on large scales. This is quantified by calculating the corresponding gradient map, $\lvert \Delta \varv_{\phi}\rvert_\mathrm{2D}/\varv_{\phi}$. The latter visualises the small local relative deviations of the azimuthal velocity field, which are less than $10^{-2}$ for radii $r\lesssim 15$~kpc. Because this measure weakly depends on the plotting resolution, we complement this study by computing the azimuthal velocity difference of every cell to the mean of its neighboring cells in three dimensions. To this end, we construct a histogram of the difference $|\Delta \varv_{\phi}|=|\varv_{\phi}-\langle \varv_{\phi}\rangle_{\mathrm{neighbours}}|$, which we normalise by $|\varv_{\phi}|$ and $|\varv_z|$, respectively (two right-hand panels in Fig.~\ref{fig:azimuthal-velocities}). Here, $\langle \varv_{\phi}\rangle_{\mathrm{neighbours}}$ denotes the mean azimuthal velocity of the neighbouring cells within twice the radius of the cell. This histogram shows that the velocity of a computational Voronoi cell does not differ significantly from the mean azimuthal velocity value of the adjacent cells. On average, the deviation is of order $\sim 3\times10^{-3}$ times smaller than the azimuthal velocity of the cell itself. Similarly, $|\Delta \varv_{\phi}|$ typically amounts to 10 per cent of $|\varv_z|$, and the azimuthal velocity difference is below the absolute value of the vertical velocity in most cells ($|\Delta \varv_{\phi}|<|\varv_z|$) except for a small subset of cells at spiral density waves. Hence, this statistically justifies our assumption underlying Eq.~\eqref{eq: t_adv}.

Note that radial CR transport via advection and anisotropic diffusion is also strongly suppressed because of the largely toroidal magnetic field configuration in the disc \citep{2013Pakmor,2016bPakmor_winds} and because circular rotation dominates the kinetic energy density \citep{Pfrommer2021}. Any residual CR fluxes not explicitly modeled in our steady-state approach need to be simulated by evolving the CR electron and proton spectra in our MHD simulations \citep{2019MNRAS.488.2235W,Winner2020,2020MNRAS.491..993G}.

\paragraph*{Electrons.}
High-energy CR electrons lose mainly their energy via radiation processes. CR electrons with Lorentz factor $\gamma_{\mathrm{e}}$ and normalised velocity $\beta_{\mathrm{e}}=\varv_{\mathrm{e}}/c$ suffer synchrotron and IC losses at rates given by \citep[see e.g.][]{1970BlumenthalGould}
\begin{align}
b_{\mathrm{syn}}&=
\frac{4}{3} \sigma_{\mathrm{T}} c \beta_{\mathrm{e}}^{2} \gamma_{\mathrm{e}}^{2}\varepsilon_{B} ,
\label{eq: b_syn(E)}
\end{align}
and
\begin{align}
b_{\mathrm{IC}}&=\frac{4}{3} \sigma_{\mathrm{T}} c \beta_{\mathrm{e}}^{2}\gamma_{\mathrm{e}}^{2}\varepsilon_{\mathrm{ph}},
\label{eq:b_IC}
\end{align}
where $\varepsilon_{B}=B^2/(8\upi)$ is the magnetic energy density, $B$ is the root-mean square magnetic field, and we assume the Thomson-limit for IC scattering, which holds if $\gamma_{\mathrm{e}}h\nu \ll m_{\mathrm{e}}c^2$, where $\nu$ is the frequency of the incoming photon. The photon energy density $\varepsilon_\mathrm{ph}$ includes photons from the CMB as well as stellar radiation. To account for the latter, we assume that the UV light emitted by young stellar populations is re-emitted in the FIR so that we are able to infer the FIR luminosity of each cell from its current SFR, where we adopt the relation obtained by \citet{1998Kennicutt}:
\begin{equation}
\frac{\mathrm{SFR}}{M_{\odot}\mathrm{\,yr^{-1}}}=\epsilon\,4.5\times10^{-44}\frac{L_{\mathrm{FIR}}}{\mathrm{erg\,s^{-1}}}=\epsilon\,1.7\times10^{-10}\frac{L_{\mathrm{FIR}}}{L_{\odot}}.\label{eq: Kennicutt 1998}
\end{equation}
The parameter $\epsilon=0.79$ follows from assuming a \citet{2003Chabrier} initial mass function \citep{2010Crain}, yielding
\begin{equation}
\frac{L_{\mathrm{FIR}}}{L_{\odot}}=7.4\times10^{9}\frac{\mathrm{SFR}}{M_{\odot}\mathrm{\,yr^{-1}}}.\label{eq: Kennicutt 1998 (2)}
\end{equation}
We assume a Planck distribution corresponding to the FIR regime, i.e. wavelengths ranging from $8-1000\,\mathrm{\umu m}$ with a typical warm dust temperature of $\sim20\,\mathrm{K}$ \citep{2000Calzetti}. The evaluation of the energy loss rate in each cell is then performed by summing over the flux arriving from all other cells $i$ with  $\mathrm{SFR}>0$ at a distance $R_i$, i.e.
\begin{equation}
\varepsilon_{\star}=\sum_{i}\frac{L_{\mathrm{FIR}}}{4\upi R_{i}^{2}c}
\label{eq: photon energy density simulations}
\end{equation}
and use $R_i=[3V_i/(4\upi)]^{1/3}$ as the distance if the considered cell is actively star forming, where $V_i$ denotes the cell's volume.
Because the computational cost of this sum would otherwise be proportional to the square of the cell number, we accelerate its computation with a tree code.
We use $\varepsilon_\star$ as the incident radiation field in Eq.~(\ref{eq:b_IC}), together with the CMB, i.e., $\varepsilon_{\mathrm{ph}}=\varepsilon_\star+\varepsilon_{\mathrm{CMB}}$. For these assumptions, the effect of the Klein-Nishina suppression of the IC emission is expected to be negligible because it is only relevant if the energy of the incoming photon becomes comparable to the electron rest mass. In the rest frame of the electron this amounts to $\gamma_{\mathrm{e}}h\nu \simeq m_{\mathrm{e}} c^2$. For IR photons ($h\nu \approx 10^{-2}\,\mathrm{eV}$), the Klein-Nishina suppression would thus only become relevant above $\gamma_{\mathrm{e}}\approx 5\times 10^7$ (or $E_{\mathrm{e}}\sim 25\,\mathrm{TeV}$). Because this is larger than the cut-off in the primary electron spectrum, Klein-Nishina effects can only become relevant for secondary electrons, which can also be produced at higher energies. Additionally, including UV radiation for the incoming photon field, with typical temperatures around $10^4\,\mathrm{K}$ (or $h\nu \approx 2.4\,\mathrm{eV}$) would become relevant for electrons with $\gamma_{\mathrm{e}}\approx 2\times 10^5$ (or $E_{\mathrm{e}}\sim 100\,\mathrm{GeV}$) and could change the detailed shape of the electron spectrum, as recently pointed out by \cite{2020bEvoli}.

Third, for losses due to bremsstrahlung emission, we assume a fully ionized medium (see Eq.~\ref{eq:b_brems}).
Besides the energy loss processes that lead to the emission of photons, Coulomb interactions with the ambient medium have to be taken into account (see Eq.~\ref{eq:b_Coulomb_e}). They typically affect the low-energy part of the electron spectrum.

\begin{figure*}

\begin{centering}
\includegraphics[]{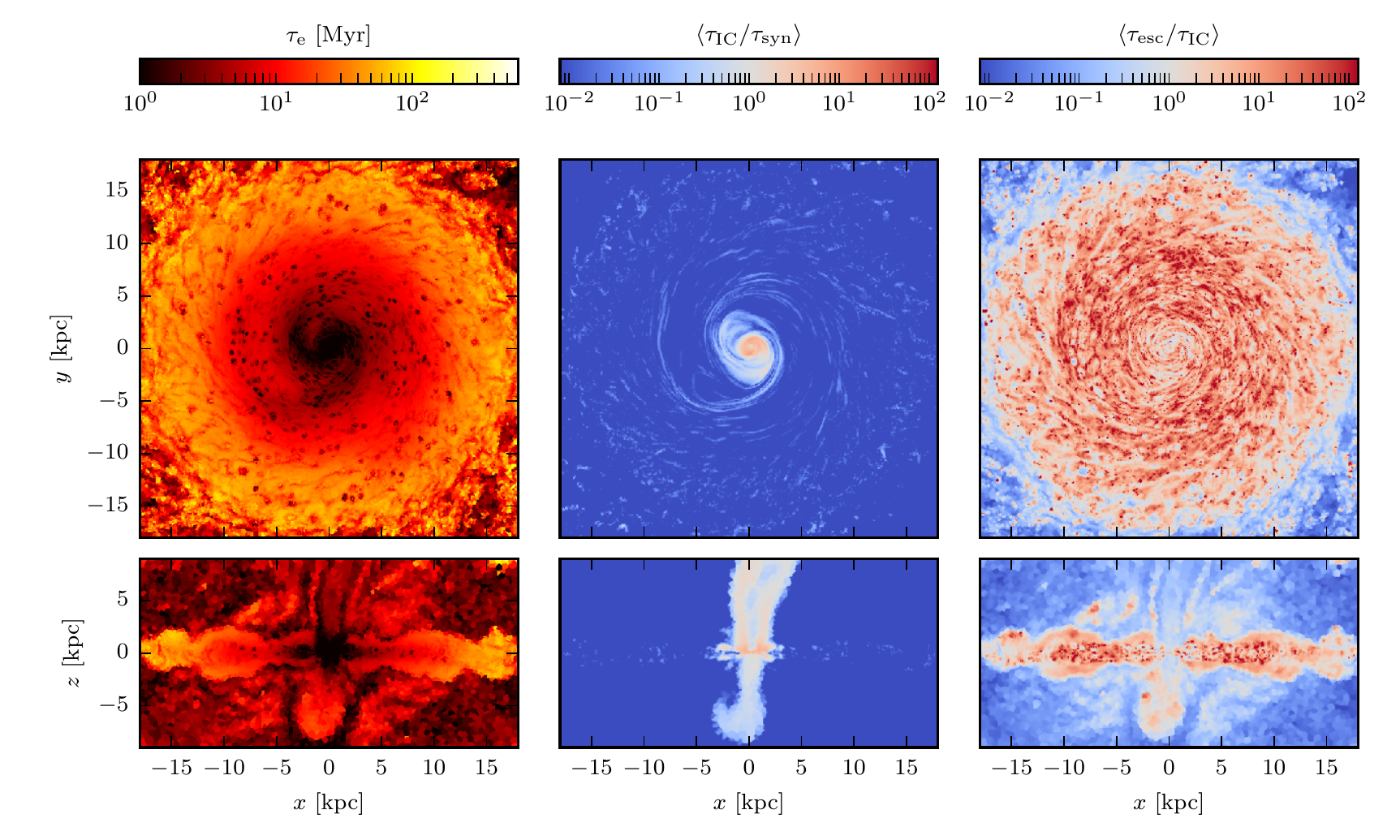}
\par\end{centering}

\begin{centering}
\includegraphics[]{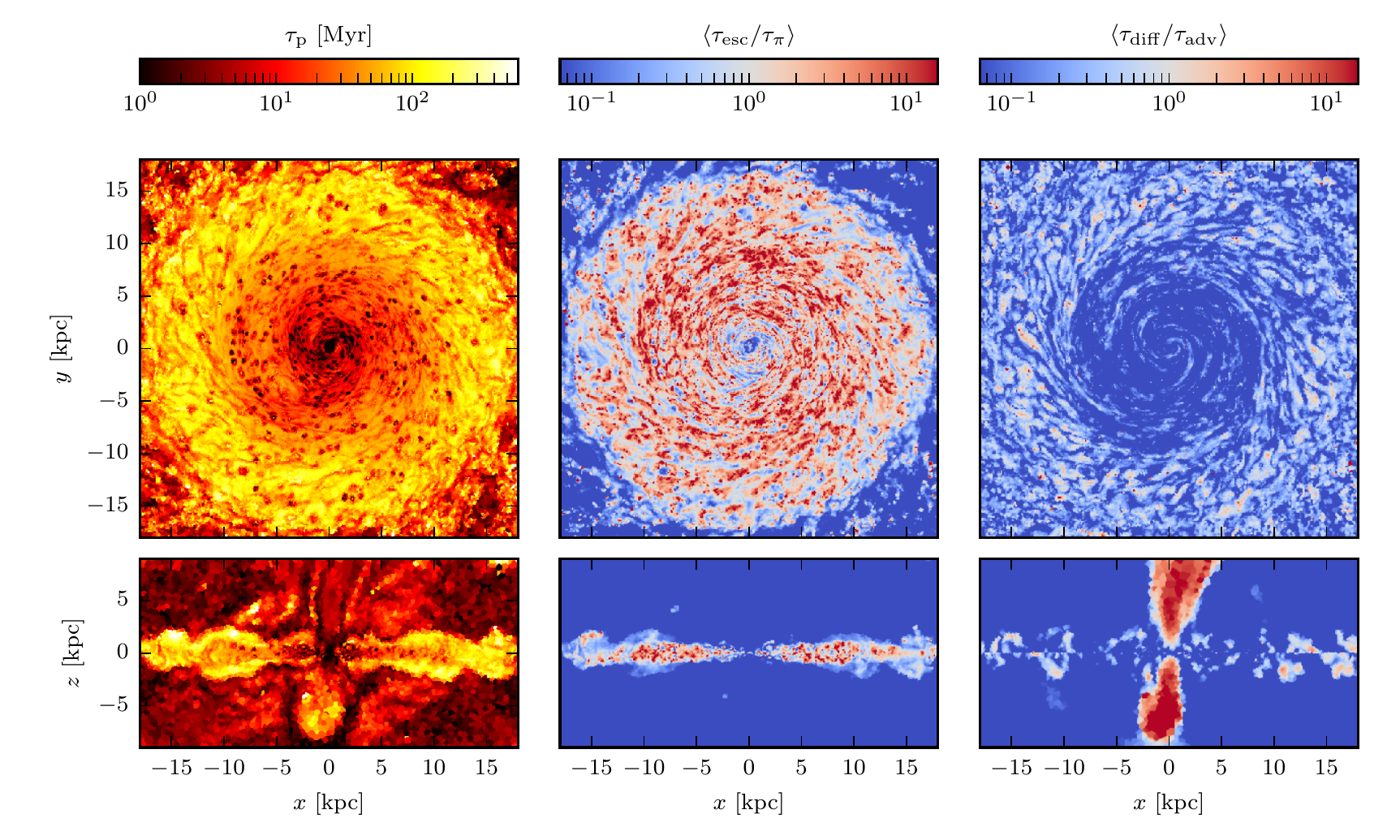}
\par\end{centering}
\caption{We show maps of characteristic timescales and their ratios at an energy of 10~GeV for our fiducial galaxy at a simulation time of 1.1~Gyr. The upper panels show the total electron cooling timescale $\tau_{\mathrm{e}}$ (left), that includes all cooling processes. The middle and right panels show the IC-to-synchrotron cooling time and escape-to-IC cooling time. The lower panels show maps of the total cooling timescale of protons $\tau_{\mathrm{p}}$ (left), the time scale ratios of escape-to-hadronic cooling (middle) and of diffusion-to-advection losses (right). All ratios are averaged over slices with a thickness of 500pc.}
\label{maps-timescales}
\end{figure*}

The discussed energy loss processes for CR electrons occur on characteristic timescales
\begin{equation}
\tau_{\mathrm{loss}}=\frac{E}{b_i(E)}\label{eq: energy loss timescale},
\end{equation}
where $b_i$ denotes the various CR electron cooling rates. They allow us to determine the importance of each energy loss process for a given energy. In Fig.~\ref{maps-timescales} we analyse our fiducial galaxy with a halo mass of $10^{12}\ M_{\odot}$ at $t = 1 $.1\,Gyr (i.e., identical to the simulations shown in Fig.~\ref{maps-properties}) and show maps of the ratio of different cooling timescales at an energy of $E=10\,\mathrm{GeV}$, averaged over a thin slice around the mid-plane of the disc with a thickness of 500 pc. The total electron cooling timescale $\tau_{\mathrm{e}}$, which includes all electron cooling processes as discussed above, is shortest in the central kpc in the disc and increases outwards up to cooling times $\tau_{\mathrm{e}}\sim 100\,\mathrm{Myr}$.
Figure~\ref{maps-timescales} shows that synchrotron losses of electrons only dominate in the very central regions, where the magnetic field is strongest. Otherwise, losses due to IC scattering occur on the shortest timescale in the disc and a few kpc above it, before escape processes take over and dominate over radiative losses. 

In the case of CR protons, the shortest timescale within the disc is the hadronic timescale because of its dependence on gas density. It is acting on typical timescales of a few tens to a few hundred Myrs. In regions of lower gas density, in the vicinity of SNRs and in outflows, escape processes are faster than hadronic interactions. Comparing the two escape losses considered here, we find that losses due to advection are predominantly occurring within the outflow, where $\varv_z$ is large and therefore $\tau_\mathrm{adv}<\tau_{\mathrm{diff}}$, while diffusion dominates elsewhere. Due to their energy dependence, we expect diffusion losses to gain importance at higher proton energies.

\subsubsection{Normalization of the CR spectra and the determination of $K_{\mathrm{ep}}^{\mathrm{inj}}$}

After solving the spectral transport equation for CR protons and electrons (Eq.~\ref{eq:N(E) from diff. loss equation}) in each Voronoi cell of our MHD simulation, we need to re-normalise the resulting steady-state spectra. The steady-state CR proton spectrum can be re-normalised to reproduce the simulated CR energy density in each cell. However, to obtain the correct normalization of the primary electron spectrum, we need a relation between the source functions of primary electrons and protons, which we can apply in each cell. To this end, we normalise the simulated electron spectrum of a Milky Way-like galaxy, averaged around the solar galactocentric radius\footnote{In practice, we average over a torus-shaped region defined by $5\,\mathrm{kpc}<r<11\,\mathrm{kpc}$ and $h<1\,\mathrm{kpc}$ at 5 Gyr. } to the observed electron-to-proton ratio at a kinetic energy of $10\,\mathrm{GeV}$, which is given by $K_{\mathrm{ep}}^{\mathrm{obs}}=10^{-2}$ \citep{Cummings2016}. We use this information to infer the corresponding injection spectrum of primary electrons, $Q_{\mathrm{e}}^{\mathrm{prim}}$, and the injection ratio $K_{\mathrm{ep}}^{\mathrm{inj}}$. 

In this way, we obtain a ratio of injected electrons to protons in each cell. By construction, the mean of this distribution averaged around the Solar circle reproduces the observed value of the electron-to-proton ratio after taking into account all cooling processes. Naturally, for a specific value of $K_{\mathrm{ep}}^{\mathrm{obs}}$, we will thus get a distribution of injection ratios, $K_{\mathrm{ep}}^{\mathrm{inj}}$. The individual steps of this procedure are explained in detail in Appendix~\ref{sec: normalization of CR spectra}.

Assuming that this injection ratio is universal, we can then apply it to the remaining part of the galaxy and to other simulated galaxies with different masses. Note that conceptually, in this framework, injection relates to effective injection $q(E)$ into a computational Voronoi cell and should not be confused with instantaneous CR injection at an individual SNR. Therefore, we do not aim to reproduce observed ratios at SNRs with our resulting value for $K_{\mathrm{ep}}^{\mathrm{inj}}$.

\begin{figure}
\begin{centering}
\includegraphics[width=\linewidth]{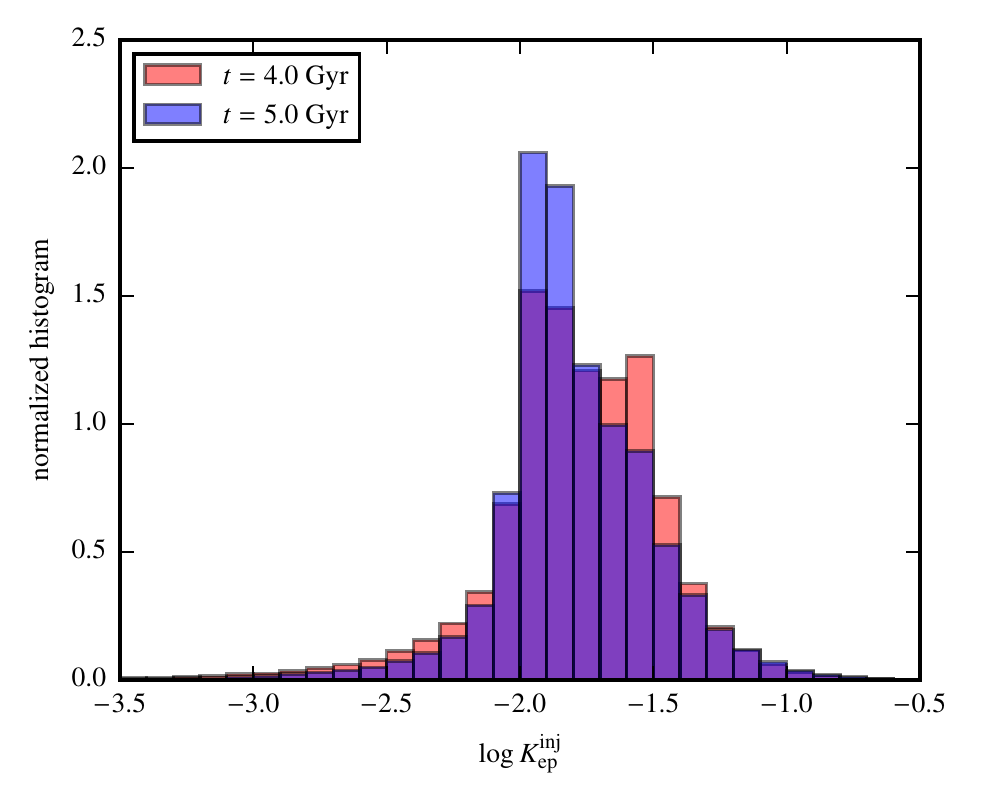}
\par\end{centering}
\caption{Histogram of the electron-to-proton injection ratio $K_{\mathrm{ep}}^{\mathrm{inj}}$ for our fiducial galaxy with a halo mass of $10^{12}\,\mathrm{M_{\odot}}$ at two times (indicated in the legend) so that the simulated SFRs at these times are both consistent with the Milky-Way value. \label{fig:Histogram-of-KepInj}}
\end{figure}

In order to quantify this dispersion of $K_{\mathrm{ep}}^{\mathrm{inj}}$ in our simulated galaxy, we use Eq.~(\ref{eq: K_ep_inj = Q_e/Q_p mp/me}), where we only consider CR spectra in a galacto-centric ring at the solar radius and re-normalise the proton injection function $q_{\mathrm{p}}$ to the CR energy density in each cell (see Eq.~\ref{eq: Q_p renormalization}) and the primary electron injection function $q_{\mathrm{e}}^{\mathrm{prim}}$ to the observed value $K_{\mathrm{ep}}^{\mathrm{obs}}$ (see Eq.~\ref{eq:Q_e normalization}). Figure~\ref{fig:Histogram-of-KepInj} shows a histogram of $K_{\mathrm{ep}}^{\mathrm{inj}}$, around the solar radius with $5\,\mathrm{kpc}<r<11\,\mathrm{kpc}$ and $h<1\,\mathrm{kpc}$, for two snapshots at 4 and 5 Gyr that exhibit global SFRs of $2.4\,\mathrm{M_{\odot}\,yr^{-1}}$  and  $1.7\,\mathrm{M_{\odot}\,yr^{-1}}$, respectively. Hence, their SFRs are comparable to the observationally inferred galactic value of $1.9\,\mathrm{M_{\odot}\,yr^{-1}}$ \citep{2011Chomiuk}. 
Despite the rather short electron cooling timescales, we obtain a narrow distribution in $K_{\mathrm{ep}}^{\mathrm{inj}}$ with mean $\langle K_{\mathrm{ep}}^{\mathrm{inj}}\rangle\approx0.02$ in both snapshots and apply this value to all evolutionary states of our simulated galaxies.\footnote{We confirmed that this result is robust to variations of these parameters: the mean $\langle K_{\mathrm{ep}}^{\mathrm{inj}}\rangle$ varies from 0.017 to 0.024 if we average over $8\,\mathrm{kpc}\pm\Delta r$ with $\Delta r$ ranging from 1 to 3 and $h$ from 0.5 to 5 kpc.} This enables us to obtain the normalization of the primary electron spectrum in other galaxies or outside the solar circle.

\begin{figure*}
\includegraphics[]{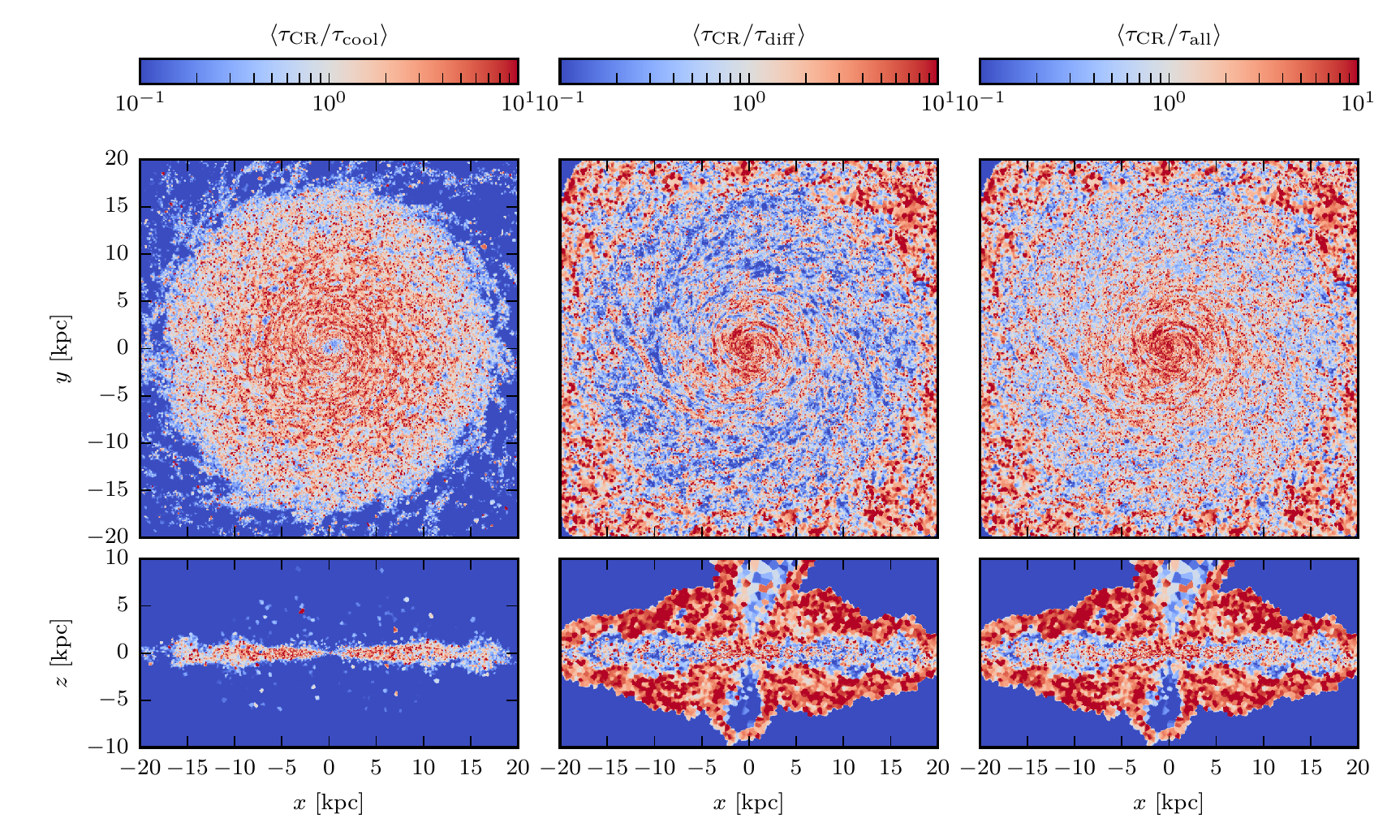}
\includegraphics[scale=1]{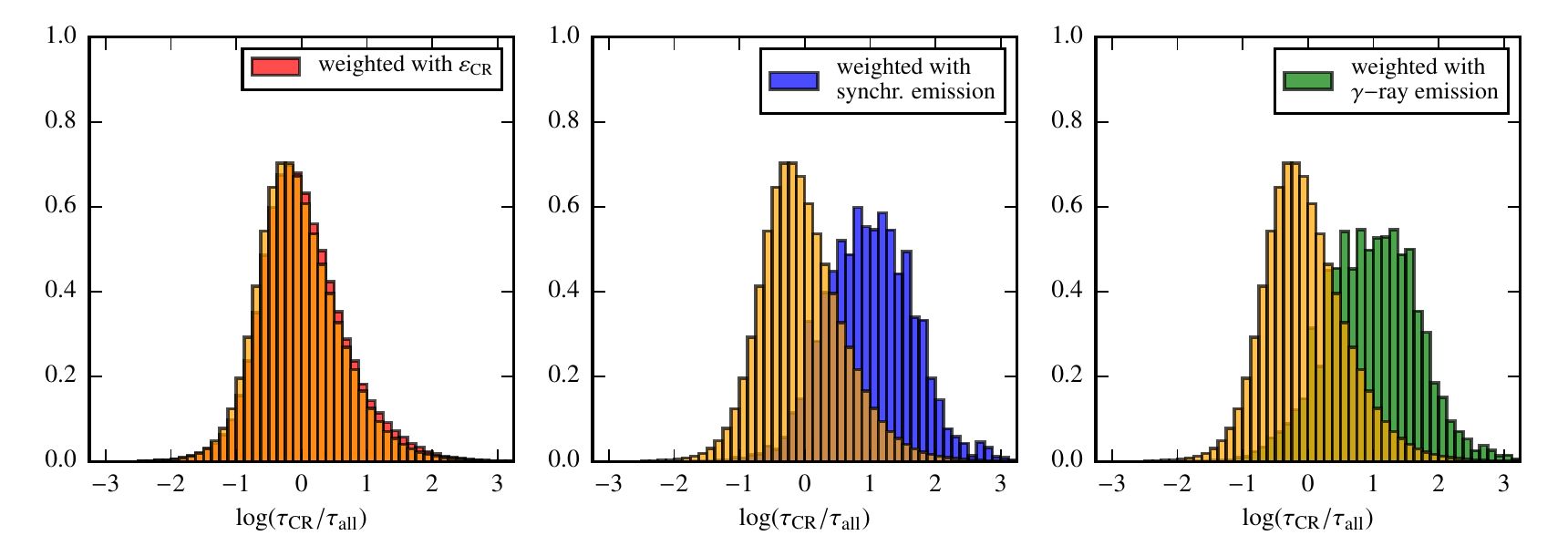}
\caption{Face-on (top three panels) and edge-on (middle three panels) projections through a thin slice (0.5 kpc) of the ratio of the timescale of the change in total CR energy density  $\tau_{\mathrm{CR}}$ to the CR cooling time (left), to the CR diffusion time (middle), and to the combined timescale of losses $\tau_{\mathrm{all}}$ (right) for our fiducial galaxy. Bottom panels: mass-weighted histograms of the cooling time ratio $\tau_{\mathrm{CR}}$/$\tau_{\mathrm{all}}$ (yellow), weighted with the CR energy density (left panel, red), the synchrotron emission (middle panel, blue) and the gamma-ray emission (right panel, green) of each cell.} \label{fig:Histogram-of-timescale-ratios}
\end{figure*}

\subsubsection{Applicability of the steady-state assumption}
In order to scrutinise our steady-state assumption and identify its caveats, we compare the change of total CR energy density in each simulation cell over a global timestep (of 0.76~Myrs) and infer a corresponding timescale $\tau_{\mathrm{CR}}=\varepsilon_{\mathrm{CR}}/ \dot{\varepsilon}_{\mathrm{CR}} $, i.e., the characteristic timescale of the change in total energy density of CRs. The purpose of $\tau_{\mathrm{CR}}$ is to provide an instantaneous time-scale (consistent with the numerical discretisation of the time integration used in the simulation) on which we report the relative change in CR proton energy density. Changes in this quantity on longer time intervals would probe CR evolution and would not represent instantaneous changes. On smaller time intervals, the hydrodynamic quantities would not represent a consistent state for Voronoi cells on the largest timesteps by construction. All cooling processes in the diffusion-loss equation should be of the same order or faster than that timescale, such that a steady state can be maintained, i.e., $\tau_{\mathrm{all}} \lesssim \tau_{\mathrm{CR}}$. Here, $\tau_{\mathrm{all}}$ is the combined rate of all relevant cooling processes at a given energy and the diffusion loss rate, i.e., $\tau_{\mathrm{all}}^{-1}= \tau_{\mathrm{cool}}^{-1}+\tau_{\mathrm{diff}}^{-1}$. As demonstrated in Fig.~\ref{maps-timescales}, the advection time-scale is larger than the diffusion time-scale with the exception of CR-driven galactic outflows. Thus, this justifies our neglect of considering the advection process in $\tau_{\mathrm{all}}$.

In Fig.~\ref{fig:Histogram-of-timescale-ratios} we show the ratio of these timescales at 10~GeV. The upper panel shows maps centered on the mid-plane of the disc, in which we average over slices with a thickness of 500~pc. We obtain  $\tau_{\mathrm{all}} \lesssim \tau_{\mathrm{CR}}$, the condition for a steady-state, predominantly inside the disc. This is owing to the short hadronic timescales in regions of high gas density in combination with large diffusive losses in the central region. Still, there remain some areas, where the steady-state assumption breaks down. This occurs either in regions of low gas density, where hadronic losses are weak, and/or in the vicinity of SN explosions, where CRs are freshly injected, leading to a sudden change in the CR energy density and disturbing the steady-state configuration. 

Nevertheless, the cells contributing predominantly to non-thermal emission processes respect the steady-state assumptions. This can be deduced from the lower panels of Fig.~\ref{fig:Histogram-of-timescale-ratios}, where we show normalised histograms of the ratios $\tau_{\mathrm{CR}}/ \tau_{\mathrm{all}}$ of all cells, weighted with the CR-energy density (left-hand panel), the synchrotron emission (middle panel) and the hadronic $\gamma$-ray emission resulting from neutral pion decay (right-hand panel), see \citetalias{2021WerhahnII} and \citetalias{2021WerhahnIII} for the description of the emission processes. Clearly, weighting the timescale ratios by the non-thermal emission of each cell, either synchrotron or hadronic gamma-ray emission, leads to a shift in the ratios $\tau_{\mathrm{CR}}/ \tau_{\mathrm{all}}$ towards higher values, indicating that the steady-state assumption is justified in cells that dominate the non-thermal emission. Yet, there is a non-negligible fraction of cells that do not obey the steady-state criterion and demand a more sophisticated treatment of the time evolution of CR spectra in three-dimensional MHD simulations, which we will examine in future work using new algorithms to follow CR electron and proton spectra \citep{2019MNRAS.488.2235W,2020MNRAS.491..993G,2020arXiv200906941O}.

\subsection{Secondary electrons and positrons}
In addition to primary CR electrons that are accelerated at sources  such as SNRs, pulsar-wind nebulae, or gamma-ray binaries, CR electrons can also be produced in inelastic collisions of CR protons with protons and other nuclei in the ambient ISM. Such hadronic reactions produce charged pions that decay into secondary electrons (and neutrinos) and the neutral pions into $\gamma$-rays.
%\begin{eqnarray}
  %\pi^\pm &\rightarrow& \mu^\pm + \nu_{\mu}/\bar{\nu}_{\mu} \rightarrow
  %e^\pm + \nu_{e}/\bar{\nu}_{e} + \nu_{\mu} + \bar{\nu}_{\mu},\nonumber\\
 % \pi^0 &\rightarrow& 2 \gamma \,.\nonumber
%\end{eqnarray}

For the calculation of the production spectrum of secondary electrons and positrons, parametrizations of the cross sections of pion production are required. In the following, we adopt the model by \citet{2018Yang} for the low energy range $T_{\mathrm{p}}<10\,\mathrm{GeV}$ (see equations \ref{eq:production of secondaries, general}, \ref{eq:dsigma_s/dE_s} and \ref{eq:dsigma/dx Yang}), the description by \citet{2006PhRvD..74c4018K} for $T_{\mathrm{p}}>100\mathrm{\,GeV}$ (see Eq.~\ref{eq: electron source function from kelner}) and a cubic spline interpolation in between. We describe the calculation of the production spectrum of secondary particles in Appendix~\ref{subsec:Secondary-Electrons-and positrons} in more detail. Furthermore, we provide our own parametrization of the total cross section of charged pion production at low proton energies in Appendix~\ref{subsec:parametrizations_for_sigma_pi}. We compare our approach to an analytical approximation in Appendix~\ref{appendix:analytical approximation q_e}, that will be useful in the following Section.

\subsection{Ratio of primary to secondary electrons}

To complement the numerical analysis of our work, we derive here an analytical approximation for the ratio of secondary electrons and positrons to primary electrons which helps to understand the physics underlying our simulation results. For simplicity, we assume that the injection spectral indices of CR electrons and protons are identical. Our analytical insight can be used to determine the relevance of each population to CR observables and quantify each contribution to the non-thermal emission, and eventually compare it to simulations.
The steady-state spectrum after taking into account all cooling losses resulting from a source function $q$ is approximately given by
\begin{equation}
f_{\mathrm{e/p}}=q_{\mathrm{e/p}}\,\tau_{\mathrm{e/p}},\label{eq:N_e/p_cooled}
\end{equation}
where $\tau_{\mathrm{e/p}}^{-1}=\tau_{\mathrm{esc}}^{-1}+\tau_{\mathrm{loss,e/p}}^{-1}$ and this equation is valid for protons, primary and secondary electrons in this analytical approximation.
Furthermore, the spectrum of secondary particles (before undergoing cooling processes) is connected to the source function of secondary particles via
\begin{align}
f_{\mathrm{e^{\pm},uncooled}}^{\mathrm{sec}}=q_{\mathrm{e^{\pm}}}^{\mathrm{sec}}\tau_{\pi},
\label{eq: f_e_uncooled = q_e*tau_pi}
\end{align}
where $\tau_{\pi}$ is the characteristic timescale of pion production or hadronic interactions of CRs with the ISM (Eq.~\ref{eq:tau_pi}). Therefore, the steady-state spectrum of secondary electrons/positrons is given by
\begin{align}
f_{\mathrm{e^{\pm}}}^{\mathrm{sec}}=q_{\mathrm{e^{\pm}}}^{\mathrm{sec}}\tau_{\mathrm{e}}=\frac{f_{\mathrm{e^{\pm},uncooled}}^{\mathrm{sec}}}{\tau_{\pi}}\tau_{\mathrm{e}}.
\end{align}
Consequently, the ratio of primary to secondary electrons can be expressed as
\begin{align}
\frac{f_{\mathrm{e}}^{\mathrm{prim}}}{f_{\mathrm{e}}^{\mathrm{sec}}}=\frac{f_{\mathrm{e}}^{\mathrm{prim}}}{2f_{\mathrm{e^{\pm},uncooled}}^{\mathrm{sec}}}\frac{\tau_{\pi}}{\tau_{\mathrm{e}}},
\label{eq:f_e,prim/f_e,sec}
\end{align}
where $f_{\mathrm{e}}^{\mathrm{sec}} = f_{\mathrm{e}^+}^{\mathrm{sec}} + f_{\mathrm{e}^-}^{\mathrm{sec}}$ is the total steady-state distribution of secondary electrons and positrons. Adopting the analytical approximation for $f_{\mathrm{e,uncooled}}^{\mathrm{sec}}/f_\rmn{p}$ (Eq.~\ref{eq:f_e,uncooled/f_p}) at a fixed physical momentum $P_0$, we get:
\begin{align}
\frac{f_{\mathrm{e,uncooled}}^{\mathrm{sec}}[P_0/(m_\mathrm{e}c)]}{f_{\mathrm{p}}[P_0/(m_\mathrm{p}c)]}\simeq\frac{128}{3} 16^{-\alpha_{\mathrm{p}}}\frac{m_{\mathrm{e}}}{m_{\mathrm{p}}}.\label{eq:N_sec/N_p analytical}
\end{align}
Combining Eqs.~\eqref{eq:N_e/p_cooled}, \eqref{eq:f_e,prim/f_e,sec}, and \eqref{eq:N_sec/N_p analytical} and evaluating the spectra at 10~GeV, such that the primary electron and proton source functions are linked by $K_{\mathrm{ep}}^{\mathrm{inj}}$ (see Eq.~\ref{eq: Q_e =K_ep_inj * Q_p}), yields
\begin{align}
\frac{f_{\mathrm{e}}^{\mathrm{prim}}}{f_{\mathrm{e}}^{\mathrm{sec}}} & =K_{\mathrm{ep}}^{\mathrm{inj}}\frac{3}{128}16^{\alpha_{\mathrm{p}}}\frac{\tau_{\pi}}{\tau_{\mathrm{p}}}\\
 & =K_{\mathrm{ep}}^{\mathrm{inj}}\frac{3}{128}16^{\alpha_{\mathrm{p}}}\left(1+\frac{\tau_{\pi}}{\tau_{\mathrm{esc}}}\right)
 \label{eq:N_e_prim/N_e_sec}\\
 &\approx0.48\,\left(1+\frac{\tau_{\pi}}{\tau_{\mathrm{esc}}}\right)
 \quad\mbox{for}\quad\alpha_{\mathrm{p}}=2.5
\end{align}
and $K_{\mathrm{ep}}^{\mathrm{inj}}\approx0.02$. Here, we assume that losses due to Coulomb interactions are negligible in comparison to hadronic losses, which is reasonable at and above the considered energies. 

\begin{figure*}
\begin{centering}
\includegraphics[]{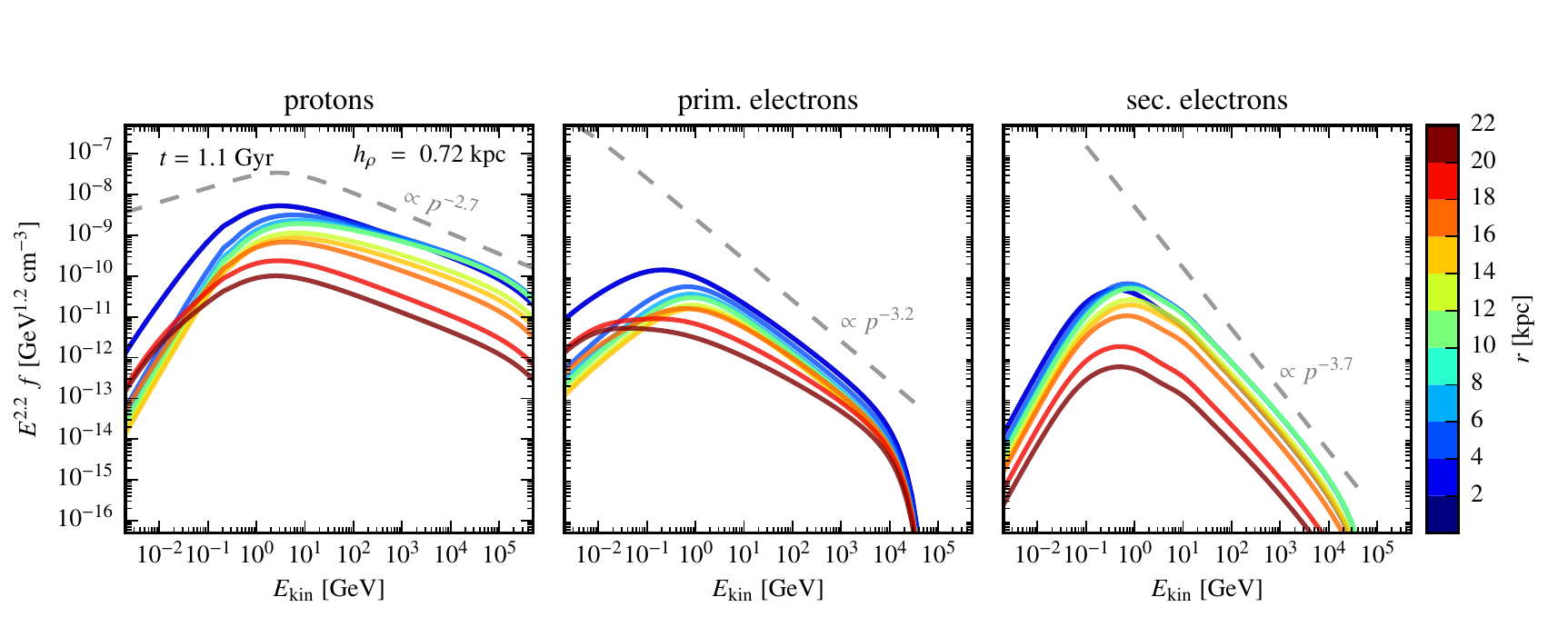}
\par\end{centering}
\caption{We show CR spectra of the different components (indicated in the titles) in radial bins as indicated by the colorbar of a simulation with halo mass $M_{200}=10^{12}\,\mathrm{M}_{\odot}$, $c_{200}=7$ and CR acceleration efficiency $\zeta_{\mathrm{SN}}=0.05$ at 1.1 Gyr.}
\label{fig:CR-spectra-radial-bins}
\end{figure*}

This implies in the fully calorimetric limit, where hadronic losses dominate over escape losses, i.e., $\tau_{\pi}\ll\tau_{\mathrm{esc}}$ (which is the case in the dense ISM, see Fig.~\ref{maps-timescales}), that the ratio of primary to secondary electrons depends only on $K_{\mathrm{ep}}^{\mathrm{inj}}$ and the spectral index of the cooled proton spectrum $\alpha_{\mathrm{p}}$. In particular, we find that in this limit, secondary electrons dominate above the primary electron population.
On the other hand, primary electrons dominate over secondary electrons as soon as escape losses are comparable to or larger than pionic losses, i.e., $\tau_{\pi}\gtrsim\tau_{\mathrm{esc}}$.
In particular, as soon as (energy dependent) diffusive losses are important and steepen the proton spectra, the resulting secondary electron spectrum will be steeper than the primary one and hence, the ratio of secondaries to primaries will decrease at higher energies.

\section{Results} \label{sec: results}
\subsection{CR spectra and maps }

In order to obtain a representative spectrum of a certain region of our simulated galaxies, we average over our cell-based steady-state spectra. In Fig.~\ref{fig:CR-spectra-radial-bins} we show the CR proton spectra of our fiducial galaxy, as well as primary and secondary electrons averaged over galacto-centric rings with different radii as indicated by the colors. The height over which the spectra are averaged is the scale height of the gas density, which in this case is $h_{\rho}=0.72\,\mathrm{kpc}$. 

We assume an injection spectral index of $\alpha_\rmn{inj}=2.2$ for protons and electrons. Energy-dependent diffusion dominates CR transport at high energies. With our assumed diffusion coefficient $D\propto E_\rmn{p}^{0.5}$, the {\it proton spectra} soften to an asymptotic spectral index of $\alpha_\rmn{p}=2.7$ (shown with a grey-dashed line in the left-hand panel of Fig.~\ref{fig:CR-spectra-radial-bins}, where the spectral change at 1~GeV is due to the relativistic dispersion relation). While this asymptotic spectrum is realised at small and large galactic radii, it is not achieved at intermediate galactocentric radii. The dominant loss process of protons at low energies is Coulomb cooling that causes the spectra to turn down in comparison to a pure momentum power-law spectrum shown with a grey-dashed line in Fig.~\ref{fig:CR-spectra-radial-bins}. This effect of Coulomb cooling below $\sim 1\,\mathrm{GeV}$ becomes more pronounced towards the denser central regions as evidenced by the stronger spectral cutoff.

While Coulomb interactions similarly affect the spectrum of {\it primary electrons} at energies $\lesssim 1$~GeV, diffusion plays a subdominant role in the central galactic regions at high energies where radiative losses due to synchrotron and IC interactions dominate. This causes the injection spectral index to steepen by one, to asymptotically arrive at the steady-state electron spectrum
\begin{align}
   f_\mathrm{e}^{\mathrm{prim}}(E_\mathrm{e})\propto \frac{1}{b_\rmn{IC} + b_\rmn{syn}}\int_0^{E_\rmn{e}}q_{\mathrm{e}}^{\mathrm{prim}}(E') \rmn{d} E'\propto E_\rmn{e}^{-(\alpha_\rmn{inj}+1)} .
   \label{eq:cooled_spectrum}
\end{align}
However, in contrast to CR protons, which only show mild spectral index variations with radius at energies larger than $\sim 10\,\mathrm{GeV}$ (see left-hand panel of Fig.~\ref{fig:CR-spectra-radial-bins}), the primary electrons undergo a change from the radiative loss-dominated regime in the centre to a more diffusion dominated regime in the outskirts of the galaxy which implies a hardening of the spectra by 0.5, yielding a spectral index of primary electrons $\alpha_{\mathrm{prim,e}}=2.7$ (see middle panel of Fig.~\ref{fig:CR-spectra-radial-bins}).

By contrast, $\alpha_{\mathrm{sec,e}}$, the spectral index of {\it secondary electrons}, is $\gtrsim \alpha_{\mathrm{prim,\,e}}$ because they originate from the steady-state CR proton population, which exhibits a steepened spectral index due to diffusive losses. After suffering radiative losses the high-energy spectral index of a steady-state secondary electron population asymptotically approaches $\alpha_{\mathrm{sec,\,e}}=3.7$, independent of galactocentric radius (right-hand panel of Fig.~\ref{fig:CR-spectra-radial-bins}).

\begin{figure*}
\begin{centering}
\includegraphics[]{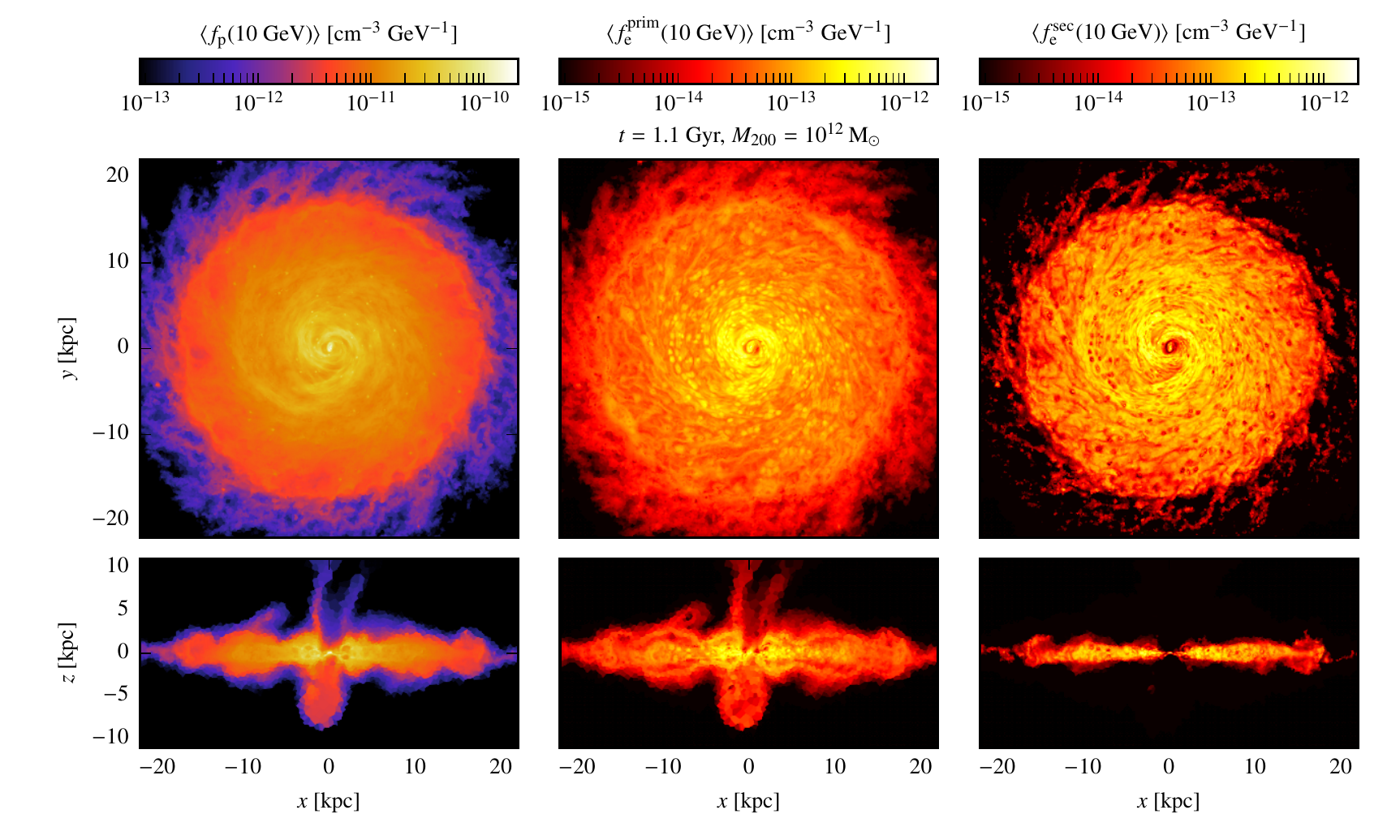}
\par\end{centering}
\begin{centering}
\includegraphics[]{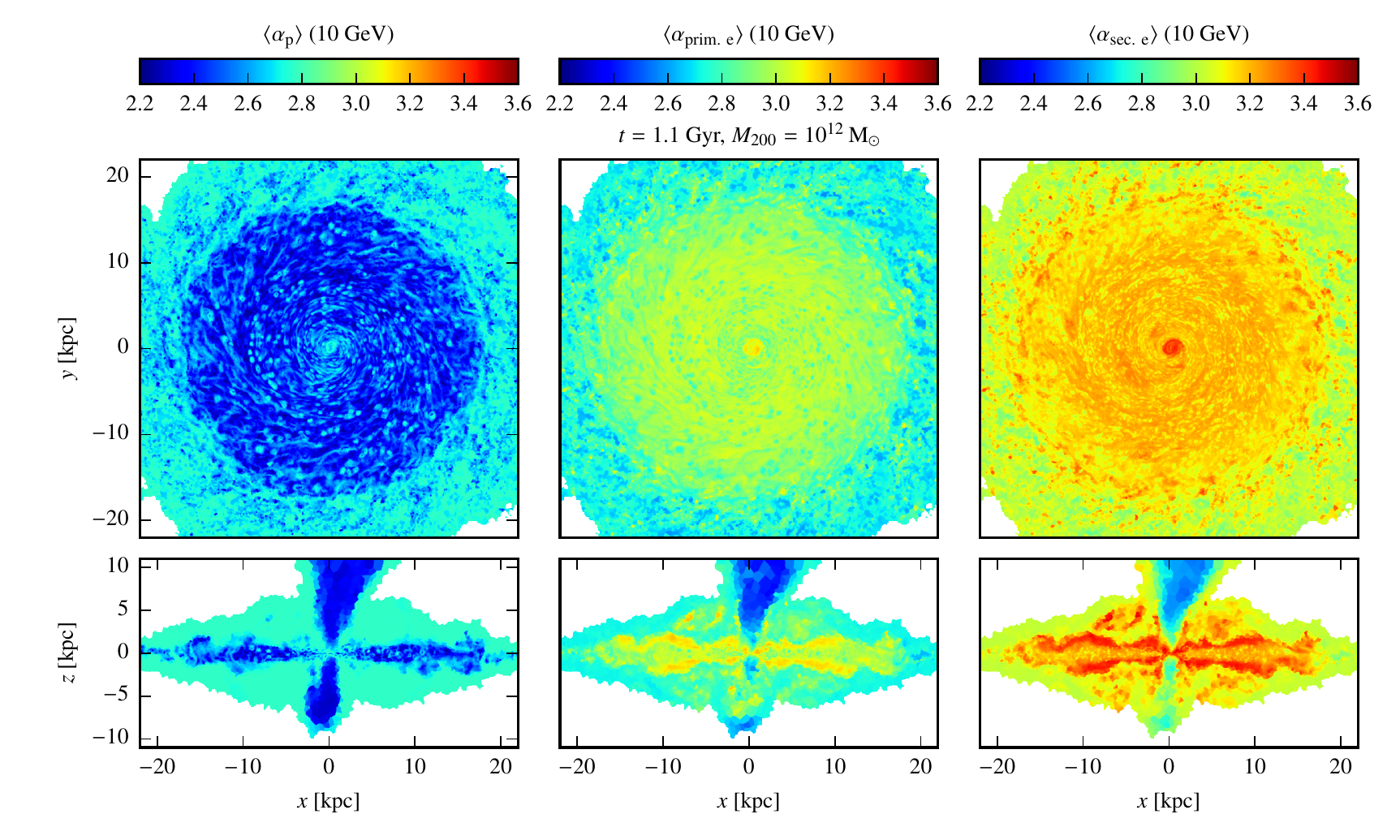}
\par\end{centering} 
\caption{We show slices of the spectral CR density at 10 GeV (top six panels for protons, primary and secondary electrons from left to right) and spectral indices (lower six panels) of CR protons ($\alpha_\mathrm{p}$), primary and secondary electrons ($\alpha_{\mathrm{prim,e}}$ and $\alpha_{\mathrm{sec,e}}$), each averaged over thin (300 pc) slices for our fiducial galaxy (see Table\,\ref{tab:simulations-overview}).}
\label{fig:CR-maps_alpha-maps}
\end{figure*}

We also show maps of the CR spectra at 10\,GeV in the upper panels of Fig.~\ref{fig:CR-maps_alpha-maps}, averaged over thin slices of 300\,pc. Similarly, the lower panels in  Fig.~\ref{fig:CR-maps_alpha-maps} show averages over 300\,pc of the spectral index at 10\,GeV of the different CR populations. We only consider cells that sum up to 99 per cent of the total energy density of the simulations in order to speed up the calculations.\footnote{This includes all cells within the disc that are relevant for our results.} Whereas the primary CR electrons reside in the same spatial regions as the CR protons, the secondary CR electron population is more concentrated towards the disc, where the gas density is large enough to yield a sufficiently large hadronic production rate. As expected, these regions coincide with those of short hadronic timescales shown in Fig.~\ref{maps-timescales}.

The spectral index analysis is performed at 10\,GeV mainly for observational reasons: (i) hadronic interactions of 10 GeV protons produce $\sim1$~GeV photons resulting from the decay of neutral pions that are well observed by \Fermi and (ii) because this is the typical electron momentum that contributes to the synchrotron radiation that is observed at 1.4\,GHz, assuming typical magnetic field strengths of $\sim 1~\mathrm{\umu G}$. Figure~\ref{fig:CR-maps_alpha-maps} shows little variation in the proton spectral index which would translate into little variation of the gamma ray spectral index provided the pion decay is the dominant gamma-ray channel. We also observe a similar contribution of primary and secondary electrons in the galactic mid-plane implying an insignificant contribution of secondaries at higher energies because of their steeper spectra. 

Moreover, secondary electrons show a more compact spatial distribution surrounding the mid-plane. Considering the more extended magnetic field distribution, we expect the primary synchrotron emission to match the secondary emission in the mid plane and to dominate over the secondary emission at higher frequencies and larger galactic heights. Similar arguments hold for the IC emission. For an incident radiation field that peaks at FIR frequency corresponding to a Planckian photon spectrum characterised by temperature 20 K, electrons with $p_{\mathrm{e}}\sim 10\,\mathrm{GeV}/(m_{\mathrm{e}} c^2)$ are able to Compton up-scatter these photons to $\sim$10\,MeV, where we expect a similar contribution of secondaries in the mid-plane, but a subdominant contribution elsewhere because the radiation field is usually also more extended than the gas distribution. 
%Those effects will be addressed in \citetalias{2021WerhahnII} and \citetalias{2021WerhahnIII} in more detail. 
The spectral index of different CR species at 10~GeV (lower panels in Fig.~\ref{fig:CR-maps_alpha-maps}) clearly show that the advection timescale dominates in the outflow regions (see also Fig.~\ref{maps-timescales}) so that the spectral index of protons and primary electrons reflects the index of the injected spectrum, $\alpha_\rmn{inj}=2.2$.

\begin{figure*}
\begin{centering}
\includegraphics[]
{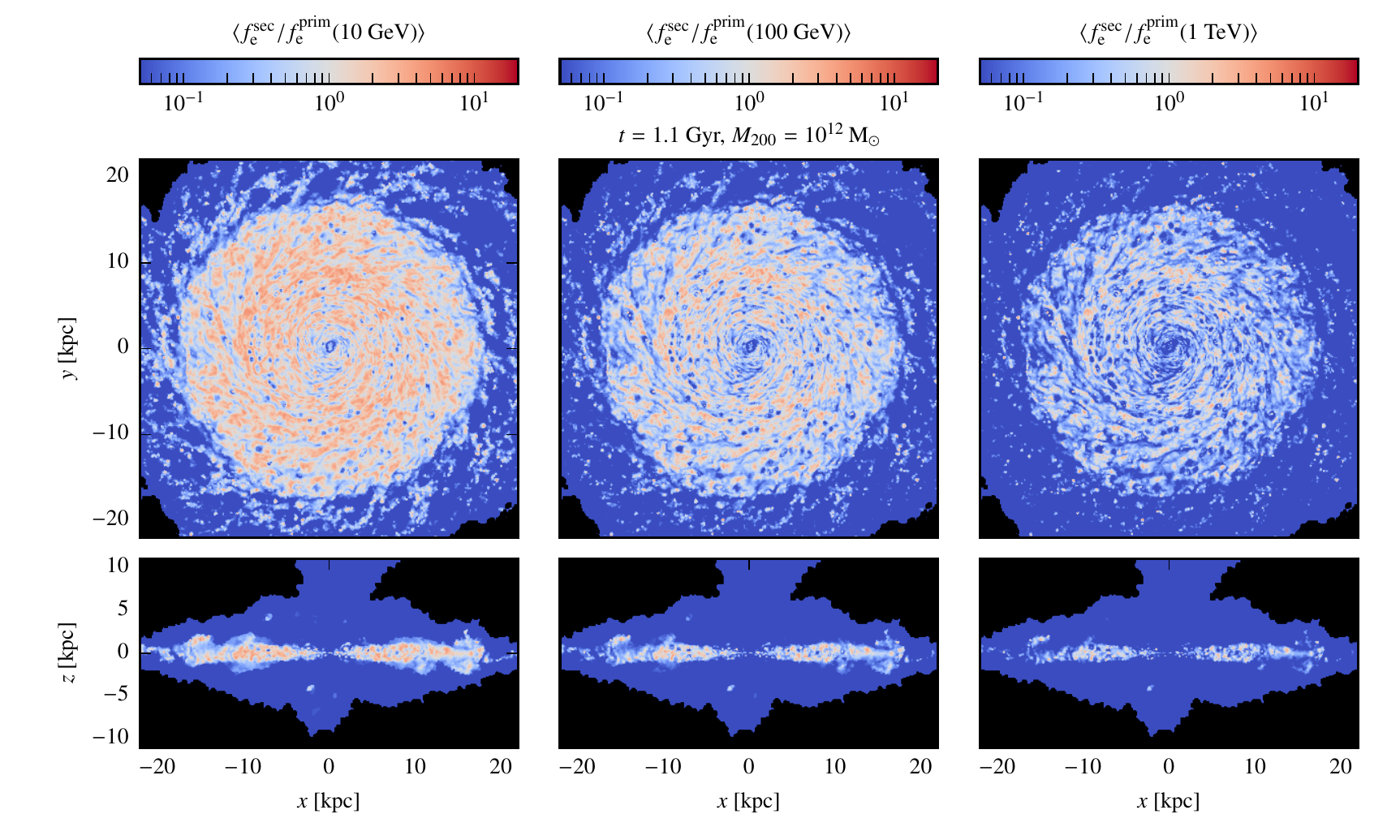}
\end{centering}
\caption{Face-on and edge-on maps of the ratio of secondary to primary electrons at 10~GeV, 100~GeV and 1~TeV averaged over a slice with thickness 300\,pc, for our fiducial galaxy, i.e. the same snapshot that is shown in Figs.\,\ref{maps-properties}, \ref{maps-timescales} and \ref{fig:CR-maps_alpha-maps}.}
\label{fig:CR-maps-sec-to-prim}
\end{figure*}

These considerations are summarised in Fig.~\ref{fig:CR-maps-sec-to-prim} that demonstrates that the ratio of primary to secondary electrons is $\gtrsim 1$ so that primaries are dominant for most parts of the galaxy. Only within the disc where the gas density is high, secondary electrons are roughly three times more abundant than primary electrons at 10~GeV. This is in accordance with Eq.~(\ref{eq:N_e_prim/N_e_sec}): using $K_{\mathrm{ep}}^{\mathrm{inj}}=0.02$, $\alpha_{\mathrm{p}}=2.2$ and assuming that $\tau_{\pi}\ll \tau_{\mathrm{esc}}$ ($\tau_{\pi}\simeq \tau_{\mathrm{esc}}$), which is required in order to produce secondaries efficiently, we obtain for the ratio in the analytical approximation  $f_{\mathrm{e}}^{\mathrm{prim}}/f_{\mathrm{e}}^{\mathrm{sec}}\approx 0.2$ ($0.4$). 
A steeper spectral index of the cooled proton spectrum $\alpha_{\mathrm{p}}>2.2$ would decrease the production rate of secondaries, which can be inferred both from our expression in Eq.~(\ref{eq:N_e_prim/N_e_sec}) and from the fact that a steeper spectral index signals a dominant role of energy dependent diffusion losses in comparison to hadronic losses, making secondary production less efficient. 
Hence, at higher energies, the importance of primary versus secondary electrons decreases further (see middle and right-hand panel of Fig.~\ref{fig:CR-maps-sec-to-prim}).

% For xi = 5%, B_0=1e-10, c=7

\subsection{Comparison to observations}

\subsubsection{Proton and electron spectra}
Since crossing the heliopause in August 2012, Voyager 1 has been observing CRs in the local interstellar medium at energies below $\sim$1~GeV, which are not subject to solar modulation effects \citep{Cummings2016}. Complementary, AMS-02 observed CR spectra at larger energies \citep{Aguilar2014a,2015Aguilar}. We see that CR proton spectra are affected by solar modulation at energies $\lesssim 10$\,GeV \citep{Potgieter_2013}.

\begin{figure*}
\begin{centering}
\includegraphics[scale=1]
{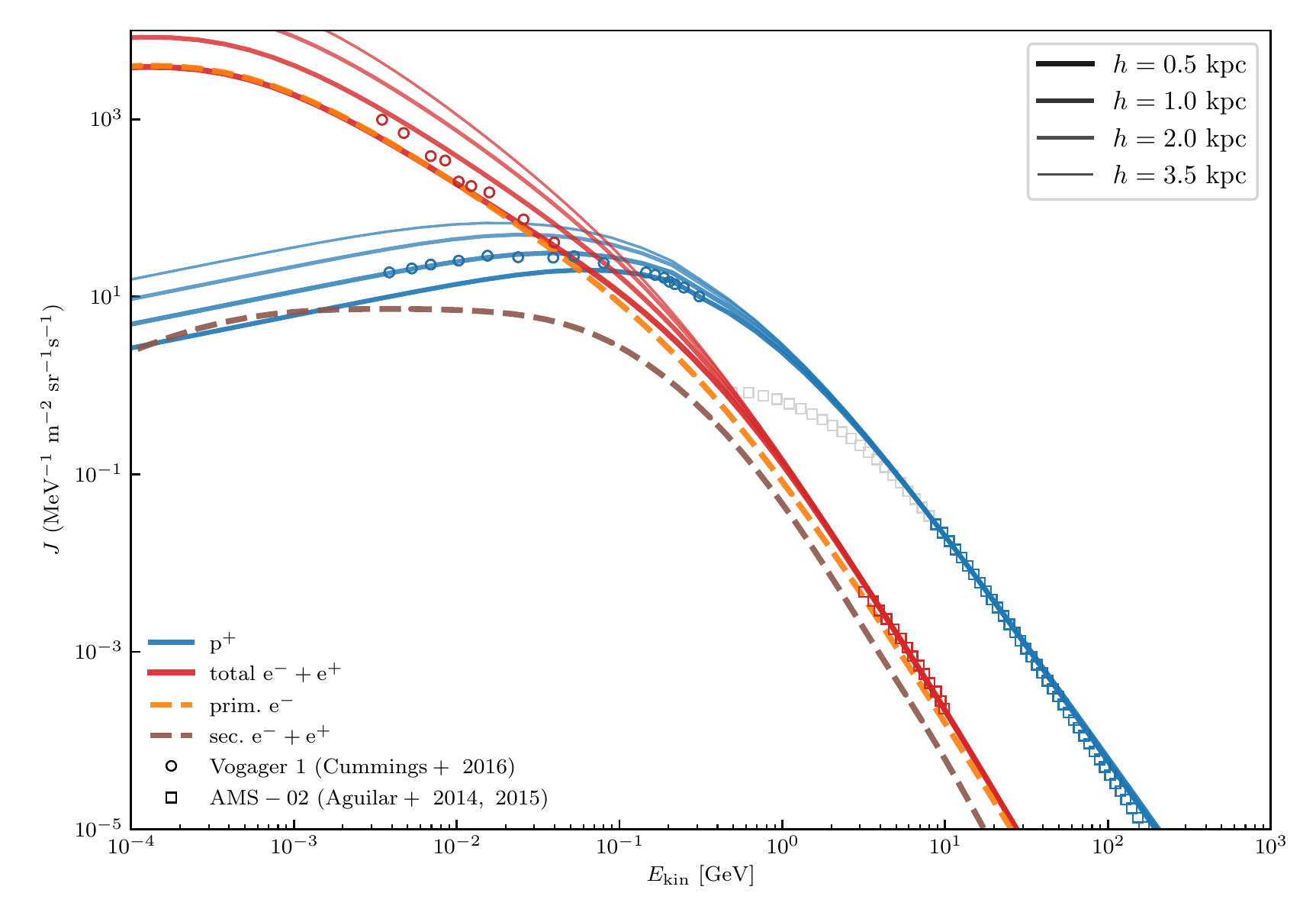}
\par\end{centering}
\caption{Spectra of CR protons and electrons at  $t=5\thinspace\mathrm{Gyr}$, averaged over rings at $r=8\pm 1 \mathrm{kpc}$ with different heights. For comparison, we show data of Voyager~1 \citep{Cummings2016} and AMS-02 for electrons \citep{Aguilar2014a} and protons \citep{2015Aguilar}. The AMS-02 data affected by solar modulation, i.e. below 10\,GeV, is shown with grey symbols. The simulated CR spectra are normalised to the observations at 10\,GeV to account for deviations of our simulated from the observed SFR (see Table \ref{tab:CR-normalization-factors} for details).}
\label{fig: CRspectrum-vs-data}
\end{figure*}

In Fig.~\ref{fig: CRspectrum-vs-data} we show both sets of observational data together with our spectra of CR protons and electrons of a Milky Way-mass galaxy ($M_{200}=10^{12}\,\rmn{M}_\odot$) at 5 Gyr, averaged over a ring at $r=8\pm 1 \mathrm{kpc}$ to mimic the conditions of the galactocentric orbit of the Sun, and different heights above and below the midplane, i.e. $\pm h$, as indicated in the plot. We normalise our spectra to the observations at 10\,GeV, which are unaffected by solar modulation (for details, see Table~\ref{tab:CR-normalization-factors}). This accounts for differences to the recent star formation history of our simulations and the Milky Way, which determines the CR injection rate and the fact that the Milky Way is under-luminous in gamma rays in comparison to the far infrared-gamma-ray relation by a factor 2.9 (using the relation of \citealt{2020Ajello}, but see also \citealt{ 2012AckermannGamma,2016RojasBravo,2017bPfrommer}). The decreasing gas density with height over the disc causes Coulomb losses to be less efficient, elevating both electron and proton spectra at low energies with larger height.

The simulated steady-state spectra of protons and electrons provide an excellent match to the observed spectra throughout the range of energies shown if we average them over heights of $\pm1$~kpc without the need of fine-tuning. In particular, the simulations nicely reproduce the observational finding that CR electrons dominate over the protons at low energies. This is due to effective Coulomb cooling at low energies, which causes a spectral flattening so that the proton-to-electron ratio approximately scales as 
\begin{align}
    \frac{f_\rmn{p}}{f_\rmn{e}}\propto \frac{b_{\mathrm{Coul,e}}}{b_{\mathrm{Coul,p}}} = \frac{A_{\mathrm{e}}}{A_{\mathrm{p}}} \frac{\beta_{\mathrm{p}}}{\beta_{\mathrm{e}}}
    \approx\frac{\beta_{\mathrm{p}}}{\beta_{\mathrm{e}}} ,
\end{align}
where we used Eqs.~\eqref{eq:cooled_spectrum}, \eqref{eq:b_Coulomb_p}, and \eqref{eq:b_Coulomb_e}, but limited the solution of the steady-state equation to Coulomb cooling only.\footnote{The deviation of the ratio $A_{\mathrm{e}}/A_{\mathrm{p}}$ from unity is below 1 per cent for the mean electron densities of interest here ($n_{\mathrm{e}}=0.02-0.2\,\mathrm{cm^{-3}}$).} While the electron and proton spectra suffer from Coulomb cooling below 1~GeV, the proton spectrum experiences an additional $\beta_\rmn{p}$ suppression at these energies (where electrons are still relativistic, i.e., $\beta_\rmn{e}\approx1$). This causes the electron spectrum to eventually dominate the total particle spectrum. 

Still, we do not simultaneously match the spectra of electrons and protons when averaged over the same height. One possible explanation for this behavior are the different spatial regions in which we observationally probe electron and proton spectra. While the shape of the proton spectrum represents the average ISM conditions in the solar radius, the {\it observed} electron spectrum exclusively probes the shocked ISM between the Sun's bow shock and the heliosheath, in which the density is increased with respect to the ISM upstream of the bow shock. As a result, Coulomb losses are stronger, which causes an additional turn-over of the electron spectrum in comparison to the model spectrum that is averaged over heights of $\pm1$~kpc. Note that while finite particle mean-free-path effects could potentially explain a small amount of solar modulation immediately upstream the heliosheath  \citep{2013Strauss}, this is unlikely to be charge dependent and cannot explain the differing CR electron and proton spectra at low energies.

Furthermore, we obtain a somewhat harder CR proton spectrum for $E\gtrsim 50$~GeV in our model, over-predicting the observed spectrum at high energies, which is probably owing to our neglect of modeling CR streaming \citep{2012Blasi, 2018Evoli}. This calls for an improved CR transport model in MHD simulations, which delivers a realistic (spatially and temporally varying) CR diffusion coefficient in the self-confinement picture \citep{2019ThomasPfrommer,2020Thomas,2021Thomas}, which would also include the effect of a reduced CR diffusion coefficient around sources \citep{2017HAWK}. This could in turn signal the excitation of powerful CR-driven plasma instabilities \citep{Shalaby2021}.

In summary, the assumptions about the injected spectral index of protons and electrons, $\alpha_{\mathrm{inj}} = 2.2$, in combination with the observationally motivated energy dependence of the diffusion coefficient, $\propto E^{0.5}$ determine our spectrum at high energies. In addition, the modelling of $K_{\mathrm{ep}}^{\mathrm{inj}}$ yields the observed electron spectrum relative to that of protons at 10~GeV, per construction. However, the behaviour of the spectra at low energies is a prediction of our model and hence provides a physical explanation for the observed inversion of the spectra below $\sim$100~MeV. 
 
\begin{table}
\caption{Summary of the normalization factors used for the CR proton and electron spectra in Fig.~\ref{fig: CRspectrum-vs-data}, respectively, averaged over different heights $h$, in order to mach the data at 10\,GeV.}
\label{tab:CR-normalization-factors}
\begin{tabular}{lcc}
\hline
$h$ [kpc] & norm. factor protons & norm. factor electrons \\
\hline
0.5   & 0.18 & 0.20\\
1.0   & 0.21 & 0.26\\
2.0   & 0.25 & 0.37\\
3.5   & 0.31 & 0.50\\
\hline
\end{tabular}
\end{table}

\subsubsection{Positron ratio \label{subsec: positron ratio}}
\begin{figure*}
\begin{centering}
\includegraphics[scale=1]{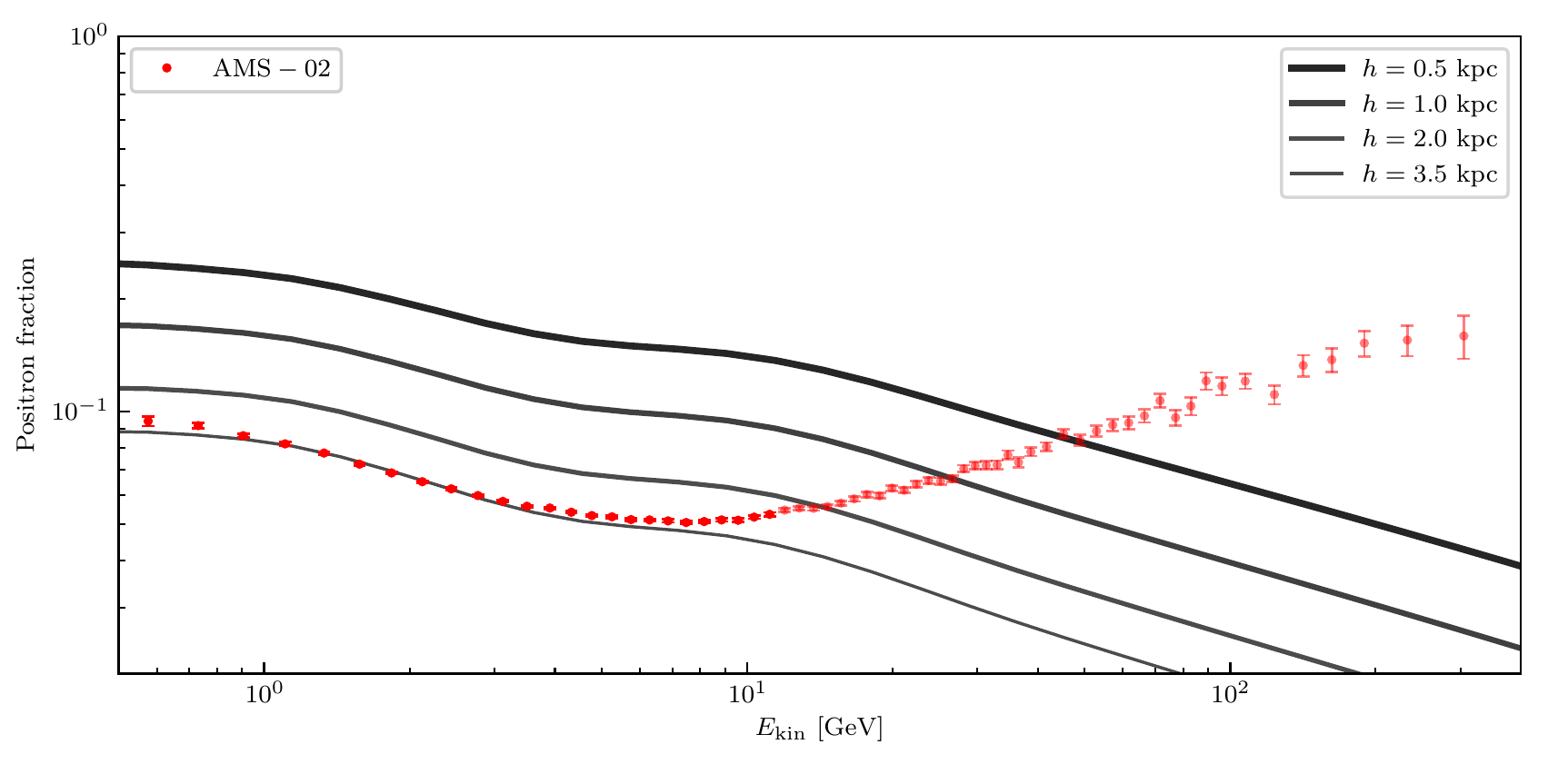}
\par\end{centering}
\caption{We show the positron fraction $f_{\mathrm{e}^{+}}/(f_{\mathrm{e}^{+}}+f_{\mathrm{e}^{-}})$, averaged over different heights, at $t=5\thinspace\mathrm{Gyr}$ (as in Fig.~\ref{fig: CRspectrum-vs-data}). The data points indicate observations performed by AMS-02 \citep{Accardo2014}. The rise of the the positron fraction above $\sim$8~GeV is due to additional positron sources that are not modelled here (see Section \ref{subsec: positron ratio}).}
\label{fig:positron-fraction}
\end{figure*}

Several experiments reported the positron ratio that decreases with energy until $\sim8$~GeV at which point it starts to rise again, including TS93 \citep{Golden_1996}, Wizard/CAPRICE \citep{BOEZIO2001}, HEAT \citep{2004Beatty}, AMS-01 \citep{AGUILAR2007}, PAMELA \citep{2009Adriani}, \Fermi \citep{Ackermann2012PosFrac}, AMS-02 \citep{Aguilar2013, Aguilar2014b}. If positrons solely arise from hadronic interactions of CR protons with the ISM, the positron fraction decreases with energy because of the steeper spectrum of secondaries in comparison to primaries. Thus, the observed rise of the positron fraction has been either attributed to annihilating/decaying dark matter particles \citep[e.g.][]{2009Yin, Cholis_2013, Feng_2020}, or local astrophysical sources such as pulsars or SNe \citep[e.g.][]{Serpico_2012, DiMauro_2017, Hooper_2009,2020Mertsch}.
We obtain in Fig.~\ref{fig:positron-fraction} the expected behaviour of a decreasing positron fraction with energy, coinciding with observations up to $\sim 8\,\mathrm{GeV}$. The overall normalization though depends on the height over which we average the CR electron spectra, as depicted in the legend. This is due to the fact that the decreasing gas density with height leads to a less efficient production of secondaries and therefore to a decrease in the fraction of positrons in comparison to all leptons, the latter being dominated by primary electrons. 

Note that we do not simultaneously reproduce the observations of CR spectra and the positron fraction within one model, when averaged over the same height. This is consistent with the fact that we somewhat overproduce the proton spectrum at energies $\sim100$ GeV in comparison to the observations (see Fig.~\ref{fig: CRspectrum-vs-data}), which produces secondary electrons at around 6~GeV. In consequence, our positron fraction tends to be higher by a factor $\approx1.9$ in comparison to the observed values. 

Another uncertainty in the calculation of the positron fraction is our assumption about the nuclear enhancement factor, that accounts for heavier nuclei in the composition of CRs and the ISM (see Appendix~\ref{subsec:Secondary-Electrons-and positrons}), for which we adopt the wounded nucleon model. However, \cite{2016Kafexhiu} recently analysed the impact of different effects such as sub-threshold pion production on the cross section of secondary particle production. In particular at low energies, the wounded nucleon model seems to break down. For example, in collisions of protons with carbon nuclei, electrons and positrons are shown to be created in equal amount, in contrast to pp-collisions, where more positrons than electrons are produced at low energies. \cite{2018Yang} apply those considerations to solar abundances and quantify the effect on the gamma-ray spectrum. Still, the exact effect on the electron and positron spectra for solar abundances has (to our knowledge) not yet been analysed.

\section{Summary and conclusions} \label{sec: summary and conclusion}
For the first time, we calculate steady state CR spectra in three-dimensionsal MHD simulations that self-consistently include CRs. We model their spectra with a cell-based steady-state approximation, including hadronic and Coulomb losses for CR protons and radiative losses due to synchrotron, bremsstrahlung and IC losses for CR electrons. In addition, we estimate losses due to advection and diffusion, while assuming an energy-dependent diffusion coefficient. Furthermore, we carefully calculate the production of secondary electrons and positrons by combining existing models from the literature \citep{2018Yang,2006PhRvD..74c4018K} at intermediate energies and provide our own parametrization of the total cross sections of negatively and positively charged pion production, respectively (Eqs.~\ref{eq: sigma-pi-minus fit formula} and \ref{eq: sigma-pi-plus fit formula}). 

Intriguingly, while our modelling of CR sources and transport in a self-consistently evolving galaxy does not involve fine-tuning of parameters, the emerging steady-state spectra convincingly reproduce observational CR features measured locally in the Milky Way. These include on the one hand measured CR proton and electron spectra in the ISM by AMS-02 \citep{Aguilar2014a,2015Aguilar} and Voyager 1 \citep{Cummings2016}, for which we reproduce the inversion of proton and electron spectra so that the latter dominates the total particle spectrum at low energies. We attribute this to steady-state Coulomb cooling and analytically verify this behavior. On the other hand, our steady-state spectra also match the observed shape of the positron fraction up to energies of $\sim8$~GeV, before the excess of positrons (attributed to additional sources such as pulsars) comes into effect. Solely the total normalization of the CR spectra and positron fraction is not reproduced. However, this is not too surprising, considering the fact that our simulations are not designed to fully reproduce a realistic Milky Way analogue with a simultaneous match of current SFR, halo mass, as well as the potentially complex star-formation history. This could be addressed in future work by applying our model to cosmological simulations, where galaxies can be found that resemble the Milky Way in more detail. Summarising, the match of simulations and observational data indicates that our modelling apparently does not miss critical physics ingredients and is able to provide physics insights into observational findings that are very complementary to state-of-the-art approaches of CR transport analyses.

Our model enables us to obtain spatial and spectral information of CRs in simulated galaxies with different galaxy sizes and injection efficiencies of CRs. We use this approach as a starting point to analyze the non-thermal emission processes arising from CRs in star-forming galaxies in two accompanying publications. \citetalias{2021WerhahnII} focuses on gamma-ray emission processes, examines the influence of CR transport models on the total gamma-ray luminosity and emission spectra, whereas \citetalias{2021WerhahnIII} is aimed towards understanding the radio emission of star-forming galaxies. 
In particular, our analytical modelling of the ratio of secondary-to-primary electrons enables us to constrain their relative contribution to the non-thermal emission processes.

\section*{Acknowledgments}
We thank our referee for a careful reading and an insightful report that helped to improve the paper. MW, CP, and PG acknowledge support by the European Research Council under ERC-CoG grant CRAGSMAN-646955. This research was supported in part by the National Science Foundation under Grant No.\ NSF PHY-1748958.

\section*{Data Availability}

The simulations and data analysis scripts underlying this article will be shared on reasonable request to the corresponding author. The \arepo code is publicly available.

\bibliographystyle{mnras}
\bibliography{literatur,literature-pg}

\appendix

\section{Loss rates and normalization of CR spectra} \label{appendix: loss rates and normalisation}

\subsection{Loss rates of CR protons and electrons}
\paragraph{Protons.}
CR protons with an energy above the threshold energy of pion production $E>E_{\mathrm{th}}=1.22\thinspace\mathrm{GeV}$ lose energy at a loss rate given by
\begin{align}
b_{\pi}=K_{\mathrm{p}}T_{\mathrm{p}}c\sigma_{\mathrm{pp}}n_{\mathrm{N}},
\label{eq: b_pi}
\end{align}
where the inelasticity of the pp-interaction is $K_{\mathrm{p}}=1/2$ \citep{1994A&A...286..983M}, $T_{\mathrm{p}}$ is the particle kinetic energy, $\sigma_{\mathrm{pp}}$ is the cross section of proton-proton collisions (given by Eq.\ \ref{eq: sigma_pp Pfrommer}) and $n_{\mathrm{N}}=n_{\mathrm{H}} + 4n_{\mathrm{He}} =(X_{\mathrm{H}} + 1- X_{\mathrm{H}})\rho/m_{\mathrm{p}}=\rho/m_{\mathrm{p}}$ is the number density of target nucleons, where $X_{\mathrm{H}}=0.76$ denotes the hydrogen fraction and $\rho$ is the gas density.

Additionally, CR protons lose energy through Coulomb interactions \citep{GOULD1972379}, given by
\begin{align}
b_{\mathrm{Coul,p}}&=\frac{3 \sigma_{\mathrm{T}} n_{\mathrm{e}}m_{\mathrm{e}}c^3 }{2\beta}\left[\ln\left(\frac{2\gamma m_{\mathrm{e}}c^{2}\beta^{2}}{\hbar\omega_{\mathrm{pl}}}\right)-\frac{\beta^{2}}{2}\right]\\
&\equiv\frac{3 \sigma_{\mathrm{T}} n_{\mathrm{e}}m_{\mathrm{e}}c^3 }{2\beta} A_{\mathrm{p}},
\label{eq:b_Coulomb_p}
\end{align}
where $n_{\mathrm{e}}=n_{\mathrm{H}} + 2n_{\mathrm{He}} = (X_{\mathrm{H}} + (1-X_{\mathrm{H}})/2)\rho/m_{\mathrm{p}}= 0.88\rho/m_{\mathrm{p}}$ is the electron number density, $\hbar$ is the reduced Planck constant, $e$ the elementary charge, $c$ the speed of light and the plasma frequency is defined as $\omega_{\mathrm{pl}}=\sqrt{4\upi e^{2}n_{\mathrm{e}}/m_{\mathrm{e}}}$. Furthermore, $\sigma_{\mathrm{T}}=8 \upi r_0^2 /3$ is the Thomson cross-section, where  $r_0=e^2/(m_{\mathrm{e}}c^2)$ denotes the classical electron radius, and $A_{\mathrm{p}}$ defines the Coulomb logarithm and the velocity correction term in the bracket. The Lorentz factor $\gamma$ and the normalised velocities $\beta=\varv/c$ without subscripts refer to protons (electrons are denoted with a subscript $\mathrm{e}$). 

\paragraph{Electrons.}
Besides losses due to IC and synchrotron emission, CR electrons lose energy due to bremsstrahlung emission. Following \citet{1970BlumenthalGould}, this yields in the case of highly relativistic electrons the expression

\begin{equation}
b_{\mathrm{brems}}=4\alpha r_{0}^{2}cn_{\mathrm{p}}\beta_{\mathrm{e}}\gamma_{\mathrm{e}}\left[\ln(2\gamma_{\mathrm{e}})-\frac{1}{3}\right]m_{\mathrm{e}}c^2.
\label{eq:b_brems}
\end{equation}
where we assume a fully ionized medium with $n_{\mathrm{p}}=0.88 \rho / m_{\mathrm{p}}$ and $\alpha$ is the fine structure constant.
Additionally, we take Coulomb losses of CR electrons into account. 
The expression for the energy loss rate has been derived by \citet{GOULD1972145} and reads
\begin{align}
b_{\mathrm{Coul,e}} & =\frac{3\sigma_{T}n_{\mathrm{e}}m_{\mathrm{e}}c^3}{2\beta_{\mathrm{e}}} \left[\ln\left(\frac{m_{\mathrm{e}}c^{2}\beta_{\mathrm{e}}\sqrt{\gamma_{\mathrm{e}}-1}}{\hbar\omega_{\mathrm{pl}}}\right)-\right.\nonumber \\
 & \left.\ln\left(2\right)\left(\frac{\beta_{\mathrm{e}}^{2}}{2}+\frac{1}{\gamma_{\mathrm{e}}}\right)+\frac{1}{2}+\left(\frac{\gamma_{\mathrm{e}}-1}{4\gamma_{\mathrm{e}}}\right)^{2}\right]\\
 & \equiv\frac{3\sigma_{T}n_{\mathrm{e}}m_{\mathrm{e}}c^3}{2\beta_{\mathrm{e}}}   A_{\mathrm{e}},
 \label{eq:b_Coulomb_e}
\end{align}
where $A_\rmn{e}$ defines the Coulomb logarithm and various correction terms in the bracket. 

\subsection{Normalization of CR spectra \label{sec: normalization of CR spectra}}

In this section, we explain the detailed procedure of obtaining the normalization of the primary electron and proton spectra, that allow us to infer an injected ratio of electrons to protons $K_{\mathrm{ep}}^{\mathrm{inj}}$.

A priori, we do not know the normalization of the injection spectra in each cell so that we first assume $C_{i,0}=1$ in Eq.~(\ref{eq: source fct. Q(p)}), for protons and electrons. 
We calculate the steady-state spectrum $f_{\mathrm{p,0}}$ that results from a given injection spectrum and all energy loss processes, and then re-normalise the source function and spectral density via the CR energy density $\varepsilon_{\mathrm{CR}}$ in every cell:
\begin{align}
f_{\mathrm{p}}(p_{\mathrm{p}})=f_{\mathrm{p,0}}(p_{\mathrm{p}})\frac{\varepsilon_{\mathrm{CR}}}{\varepsilon_{\mathrm{CR,0}}}
\end{align}
and
\begin{align}
q_{\mathrm{p}}(p_{\mathrm{p}})=q_{\mathrm{p,0}}(p_{\mathrm{p}})\frac{\varepsilon_{\mathrm{CR}}}{\varepsilon_{\mathrm{CR,0}}}, \label{eq: Q_p renormalization}
\end{align}
where
\begin{align}
\varepsilon_{\mathrm{CR,0}}=\intop T_{\mathrm{p}}(p_{\mathrm{p}})f_{\mathrm{p,0}}(p_{\mathrm{p}})\mathrm{d}p_{\mathrm{p}}
\end{align}
and $T_{\mathrm{p}}(p_{\mathrm{p}})=\left(\sqrt{p_{\mathrm{p}}^{2}+1}-1\right) m_\rmn{p} c^2$. Similarly, for the primary electrons we calculate a steady-state spectrum $f_{\mathrm{e,0}}^{\mathrm{prim}}(p_{\mathrm{e}})$ from the injection spectrum $q_{\mathrm{e,0}}^{\mathrm{prim}}(p_{\mathrm{e}})$, and re-normalise it as
\begin{align}
f_{\mathrm{e}}^{\mathrm{prim}}(p_{\mathrm{e}}) \mathrm{d}p_{\mathrm{e}} = A_{\mathrm{norm}} f_{\mathrm{e,0}}^{\mathrm{prim}}(p_{\mathrm{e}}) \mathrm{d}p_{\mathrm{e}}.
\end{align}
To identify $A_{\mathrm{norm}}$, we relate the steady state spectra of all electrons (primary plus secondary) to protons via the observed ratio of electrons to protons $K_{\mathrm{ep}}^{\mathrm{obs}}$ at a kinetic energy of $10\,\mathrm{GeV}$, or equivalently at the corresponding normalised momenta, $p_{i\mathrm{,10\mathrm{GeV}}}$, i.e.
\begin{align}
f_{\mathrm{e}}^{\mathrm{prim+sec}}(p_{\mathrm{e,10\mathrm{GeV}}}) \mathrm{d}p_{\mathrm{e}} =K_{\mathrm{ep}}^{\mathrm{obs}}f_{\mathrm{p}}(p_{\mathrm{p,10\mathrm{GeV}}})\mathrm{d}p_{\mathrm{p}}.
\end{align}
Because we have already normalised the proton spectrum $N_{\mathrm{p}}$, we are able to obtain the primary electron spectrum via
\begin{align}
f_{\mathrm{e}}^{\mathrm{prim}}(p_{\mathrm{e}})&= A_\mathrm{norm}  f_{\mathrm{e,0}}^{\mathrm{prim}}(p_{\mathrm{e}}),\\
A_\mathrm{norm} &= \frac{\displaystyle K_{\mathrm{ep}}^{\mathrm{obs}} f_{\mathrm{p}}(p_{\mathrm{p,10\mathrm{GeV}}})\frac{ m_{\mathrm{e}}}{m_{\mathrm{p}}}-f_{\mathrm{e}}^{\mathrm{sec}}(p_{\mathrm{p,10\mathrm{GeV}}})}{f_{\mathrm{e,0}}^{\mathrm{prim}}(p_{\mathrm{e,10\mathrm{GeV}}})},
\label{eq:f_norm}
\end{align}
where we account for the normalised momenta $\mathrm{d}p_{\mathrm{p}}=\mathrm{d}p_{\mathrm{e}}m_{\mathrm{e}}/m_{\mathrm{p}}$ and use the fact that the normalised and un-normalised primary electron spectra are self-similar. Equivalently, we determine the normalization of the electron injection spectrum $q_{\mathrm{e}}^{\mathrm{prim}}(p_{\mathrm{e}})$, that is linearly related to  $f_{\mathrm{e}}^{\mathrm{prim}}(p_{\mathrm{e}})$:
\begin{align}
q_{\mathrm{e}}^{\mathrm{prim}}(p_{\mathrm{e}})= A_\mathrm{norm}  q_{\mathrm{e,0}}^{\mathrm{prim}}(p_{\mathrm{e}}),
\label{eq:Q_e normalization}
\end{align}
In order to infer the ratio of injected electrons to protons $K_{\mathrm{ep}}^{\mathrm{inj}}$, we compare the electron and proton injection spectrum at the same (physical) momentum $P_{0}$:
\begin{align}
q_{\mathrm{e}}^{\mathrm{prim}}[P_{0}/(m_\mathrm{e} c)]\mathrm{d}p_{\mathrm{e}}=K_{\mathrm{ep}}^{\mathrm{inj}}q_{\mathrm{p}}[P_{0}/(m_\mathrm{p} c)]\mathrm{d}p_{\mathrm{p}},
\label{eq: Q_e =K_ep_inj * Q_p}
\end{align}
which yields after inserting our definition of the source functions the following expression:
\begin{align}
K_{\mathrm{ep}}^{\mathrm{inj}}=\frac{q_{\mathrm{e}}^{\mathrm{prim}}[P_{0}/(m_\mathrm{e} c)]}{q_{\mathrm{p}}[P_{0}/(m_\mathrm{p} c)]}\frac{m_{\mathrm{p}}}{m_{\mathrm{e}}} =\frac{C_{\mathrm{e}}}{C_{\mathrm{p}}}\left(\frac{m_{\mathrm{p}}}{m_{\mathrm{e}}}\right)^{1-\alpha_{\mathrm{p}}}.
\label{eq: K_ep_inj = Q_e/Q_p mp/me}
\end{align}

In this way, we obtain in each cell a ratio of injected electrons to protons that eventually reproduces the observed value after taking into account all cooling processes. 
If the electrons and protons cool on the same timescales (at the considered energy of 10 GeV), we obtain $K_{\mathrm{ep}}^{\mathrm{inj}}\simeq K_{\mathrm{ep}}^{\mathrm{obs}}$. On the other hand, $K_{\mathrm{ep}}^{\mathrm{inj}}<K_{\mathrm{ep}}^{\mathrm{obs}}$ ($K_{\mathrm{ep}}^{\mathrm{inj}}>K_{\mathrm{ep}}^{\mathrm{obs}}$) implies that the timescales of the hadronic cooling processes are smaller (larger) than the leptonic ones.

In the literature, the normalization is often defined differently, in terms of injected energy into CR protons and electrons, i.e., $\varepsilon_{\mathrm{e}}^{\mathrm{inj}}= \zeta_{\mathrm{prim}}\varepsilon_{\mathrm{p}}^{\mathrm{inj}}$, such that
\begin{align}
\zeta_{\mathrm{prim}} \intop_0^\infty T_{\mathrm{p}}(p_{\mathrm{p}})q_{\mathrm{p}}(p_{\mathrm{p}})\mathrm{d}p_{\mathrm{p}}=\intop_0^\infty T_{\mathrm{e}}(p_{\mathrm{e}})q_{\mathrm{e}}^{\mathrm{prim}}(p_{\mathrm{e}})\mathrm{d}p_{\mathrm{e}}.
\end{align}
Assuming the same injected spectral index of electrons and protons, $2<\alpha_\rmn{inj}<3$, and a lower momentum cutoff that is much smaller than $m_\rmn{p}c$ ($m_\rmn{e}c$) for protons (electrons), it can be shown that this is related to $K_{\mathrm{ep}}^{\mathrm{inj}}$ in the following way:
\begin{align}
K_{\mathrm{ep}}^{\mathrm{inj}}=\zeta_{\mathrm{prim}}\left(\frac{m_{\mathrm{p}}}{m_{\mathrm{e}}}\right)^{2-\alpha_{\mathrm{p}}}%\equiv\tilde{\delta}^{-1}.
\label{eq:K_ep-Delta}
\end{align}
This implies that our approach, where we find $K_{\mathrm{ep}}^{\mathrm{inj}}\approx0.02$, this corresponds to a primary electron-to-proton energy fraction of $\zeta_{\mathrm{prim}}\approx 9\%$ (assuming $\alpha_{\mathrm{p}}=2.2$), which is consistent with the parameters used in other models, such as the one-zone steady-state models by \cite{2010Lacki}.

\section{Electron source function and parametrization of the pion cross section} \label{appendix: electron source fct. and parametrizations}

Here, we describe numerical algorithms for computing the electron source function and charged pion cross section in Appendix~\ref{subsec:Secondary-Electrons-and positrons}, before we detail our parametrization of the total cross section of pion production, $\sigma_\pi$, in Appendix~\ref{subsec:parametrizations_for_sigma_pi}. Finally, we derive an analytical approximation for the secondary electron source function in Appendix~\ref{appendix:analytical approximation q_e}.

\subsection{Production of Secondary Electrons and Positrons\label{subsec:Secondary-Electrons-and positrons}}

The minimum total proton energy required to produce a pion is $E_{\mathrm{p}}^{\mathrm{min}} = 1.22\,\mathrm{GeV}$. 
The production spectrum, i.e., the number of produced secondary particles per energy, time and volume, or source function $q_s$, of a secondary particle species $s=\gamma,\mathrm{e}^{-},\mathrm{e}^{+}$ for a given CR proton distribution is given by
\begin{align}
q_{s}(E_{s}) = c n_{\mathrm{H}} \intop_{E_{\mathrm{p}}^{\mathrm{min}}}^{\infty} \mathrm{d}E_{\mathrm{p}} 
f_{\mathrm{p}} (E_{\mathrm{p}} ) \frac{\mathrm{d}\sigma_{s}(E_{s},E_{\mathrm{p}})}{\mathrm{d}E_{s}}.
\label{eq:production of secondaries, general}
\end{align}
The differential cross section of the secondary particle species $s$ can be calculated by means of the differential cross section for the production of a pion with energy $E_{\pi}$ from the collision of a proton with energy $E_{\mathrm{p}}$, i.e. $\mathrm{d}\sigma(E_{\mathrm{p}},E_{\pi})/\mathrm{d}E_{\pi}$. Then, we can solve the integral
\begin{align}
\frac{\mathrm{d}\sigma_{s}(E_{s},E_{\mathrm{p}})}{\mathrm{d}E_{s}} = \intop_{E_{\pi}^{\mathrm{min}}}^{E_{\pi}^{\mathrm{max}}}\mathrm{d}E_{\pi}\frac{\mathrm{d}\sigma_{\pi}(E_{\mathrm{p}},E_{\pi})}{\mathrm{d}E_{\pi}}f_{s,\pi}(E_{s},E_{\pi}),
\label{eq:dsigma_s/dE_s}
\end{align}
where $f_{s,\pi}(E_{s},E_{\pi})$ is the normalised probability distribution for the production of a secondary particle $s$ from a single pion energy $E_{\pi}$. Here, we are interested in electrons and positrons ($s=\mathrm{e}^{-},\mathrm{e}^{+}$), and thus only in charged pions, while we will consider the production of neutral pions and gamma rays in \citetalias{2021WerhahnII}. 
For the normalised electron/positron energy distribution $f_{\pi^{\pm}}(E_{\mathrm{e}^{\pm}},E_{\pi})$, we use the expressions derived by \citep{1986ApJ...307...47D}, assuming a mono-energetic, unpolarized, isotropic distribution of
pions, to eventually calculate $\mathrm{d}\sigma_{\mathrm{e}^{\pm}}(E_{\mathrm{e}^{\pm}},E_{\mathrm{p}})/\mathrm{d}E_{\mathrm{e}^{\pm}}$ from Eq.\ (\ref{eq:dsigma_s/dE_s}).
%the limits in the integral of Eq.~(\ref{eq:dsigma_s/dE_s}) are given by $E_{\pi}^{\mathrm{max}}=\infty$ and 
%
%\begin{equation}
%E_{\pi}^{\mathrm{min}}=
%\left\{ 
%\begin{aligned}
%& m_{\pi}c^{2}
%& \mbox{for } E_{\mathrm{e}}<E_{\mathrm{e}}^{\mathrm{max}},\\
%& \frac{1}{2}\,m_{\pi}c^{2}
%\left(\frac{E_{\mathrm{e}}}{E_{\mathrm{e}}^{\mathrm{max}}}+\frac{E_{\mathrm{e}}^{\mathrm{max}}}{E_{\mathrm{e}}}\right)
%& \mbox{for } E_{\mathrm{e}}>E_{\mathrm{e}}^{\mathrm{max}},
%\end{aligned}
%\right.
%\label{eq:MF_high-z}
%\end{equation}
%where $E_{\mathrm{e}}^{\mathrm{max}}=m_{\mu}c^{2}(1+\beta_{\mu})\gamma_{\mu}/2\approx69.9\,\mathrm{MeV}$. 
In the literature, there are different parametrizations for the corresponding terms of Eq.~(\ref{eq:production of secondaries, general}). They relate to the definition of the pion source function in the following way
\begin{align}
q_{\pi}(E_{\pi})=cn_{\mathrm{H}}\intop_{E_{\mathrm{p}}^{\mathrm{min}}}^{\infty}\mathrm{d}E_{\mathrm{p}} f_{\mathrm{p}} (E_{\mathrm{p}} )  \frac{\mathrm{d}\sigma_{\pi}(E_{\mathrm{p}},E_{\pi})}{\mathrm{d}E_{\pi}},
\label{eq: pion-source fct. def.}
\end{align}
so that the source function of a secondary particle species is
\begin{align}
q_{s}=\intop_{E_{\pi}^{\mathrm{min}}}^{E_{\pi}^{\mathrm{max}}}\mathrm{d}E_{\pi}q_{\pi}(E_{\pi})f_{\pi}(E_{s},E_{\pi}).
\label{eq: source function for secondaries, 1}
\end{align}

\begin{figure}
\includegraphics[scale=1]{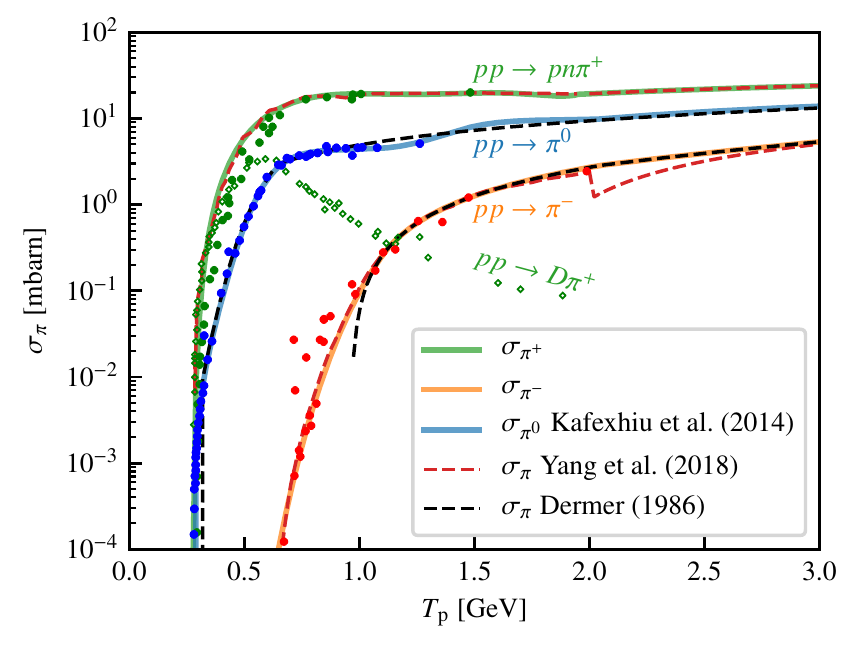}
\caption{Total cross sections of $\pi^{0}$ and $\pi^{\pm}$ production with data
points taken from the compilation in \citet{2018Yang}. The solid lines represent the parametrizations that are used here where we provide our own formulas for $\sigma_{\pi^+}$ (green) and $\sigma_{\pi^-}$ (orange) in Eqs.~\eqref{eq: sigma-pi-minus fit formula} and \eqref{eq: sigma-pi-plus fit formula}.
\label{fig:Total-cross-sections}}
\end{figure}

The differential cross section of charged pion production or the electron source function can be obtained from simulations of pp-interactions, e.g. Pythia \citep{2006_PYTHIA,2008_PYTHIA}, SIBYLL \citep{1994Fletcher_SIBYLL}, QGSJET \citep{1993_QGSJET,1997_QGSJET,2006_QGSJET} and Geant4 \citep{2003Agostinelli_Geant4,2006Allison_Geant4}. 
At low proton energies near the kinematic threshold ($T_{\mathrm{p}}<10\,\mathrm{GeV}$), we adapt the approach given by \citet{2018Yang}. They utilised the hadronic interaction model of the Geant4 Toolkit \citep{2003Agostinelli_Geant4,2006Allison_Geant4} to provide a parametrization for the normalised pion energy distribution $\tilde{f}(x,T_{\mathrm{p}})$, that reads
\begin{equation}
\frac{\mathrm{d}\sigma_{\pi}}{\mathrm{d}x}=\sigma_{\pi}\times \tilde{f}(x,T_{\mathrm{p}}).
\label{eq:dsigma/dx Yang}
\end{equation}
\citet{2018Yang} provide analytical formulae for $\tilde{f}(x,T_{\mathrm{p}})$, with $x=T_{\pi}/T_{\pi}^{\mathrm{max}}$.
The parametrization of the total cross section of pion production $\sigma_{\pi}$ is described in Section~\ref{subsec:parametrizations_for_sigma_pi}, where we provide our own fit to the data, shown by the solid lines in Fig.~\ref{fig:Total-cross-sections} and compare it to models in the literature.

In the high-energy range of protons with $T_{\mathrm{p}}>100\mathrm{\,GeV}$, we use an analytical parametrization provided by \citet{2006PhRvD..74c4018K}, where they give production rate of secondary electrons from the SIBYLL code \citep{1994Fletcher_SIBYLL}.
It is given in terms of a distribution $F_{\mathrm{e}^{\pm}}(x,E_{\mathrm{p}})$, such that $F_{\mathrm{e}^{\pm}}(x,E_{\mathrm{p}})\mathrm{d}x$ describes the number of produced electrons and positrons per collision in the interval $(x,x+\rmn{d}x)$ with $x=E_{\mathrm{e}^{\pm}}/E_{\mathrm{p}}$. Assuming that the production of secondary positrons is equal to the production of secondary electrons in this energy regime, the corresponding production rate is given by
\begin{equation}
q_{\mathrm{e}^{\pm}}(E_{\mathrm{e}^{\pm}})=cn_{\mathrm{H}}\intop_{E_{\mathrm{e}^{\pm}}}^{\infty}\sigma_\rmn{pp}^{\mathrm{inel}}(E_{\mathrm{p}}) f_{\mathrm{p}}(E_{\mathrm{p}}) F_{\mathrm{e}^{\pm}}\left(\frac{E_{\mathrm{e}^{\pm}}}{E_{\mathrm{p}}},E_{\mathrm{p}}\right)\frac{\mathrm{d}E_{\mathrm{p}}}{E_{\mathrm{p}}}.\label{eq: electron source function from kelner}
\end{equation}
%Thus, the differential cross section of secondary electron/positron
%production can be identified from their formalism to be
%\begin{align}
%\frac{\mathrm{d}\sigma_{\mathrm{e}}(E_{\mathrm{e}},E_{p})}{\mathrm{d}E_{\mathrm{e}}}=\sigma_{\mathrm{inel}}(E_{\mathrm{p}}) F_{\mathrm{e}}\left(E_{\mathrm{e}}/E_{\mathrm{p}},E_{\mathrm{p}}\right)/E_{\mathrm{p}}.
%\end{align}
Here, $\sigma_\rmn{pp}^{\mathrm{inel}}$ is the total inelastic cross section parametrized by \citet{2014PhRvD..90l3014K} (see Eq.~\ref{eq:sigma_pp_inel Kafexhiu 2014}).

\begin{figure}
%\begin{centering}
\includegraphics[width=8.4cm]{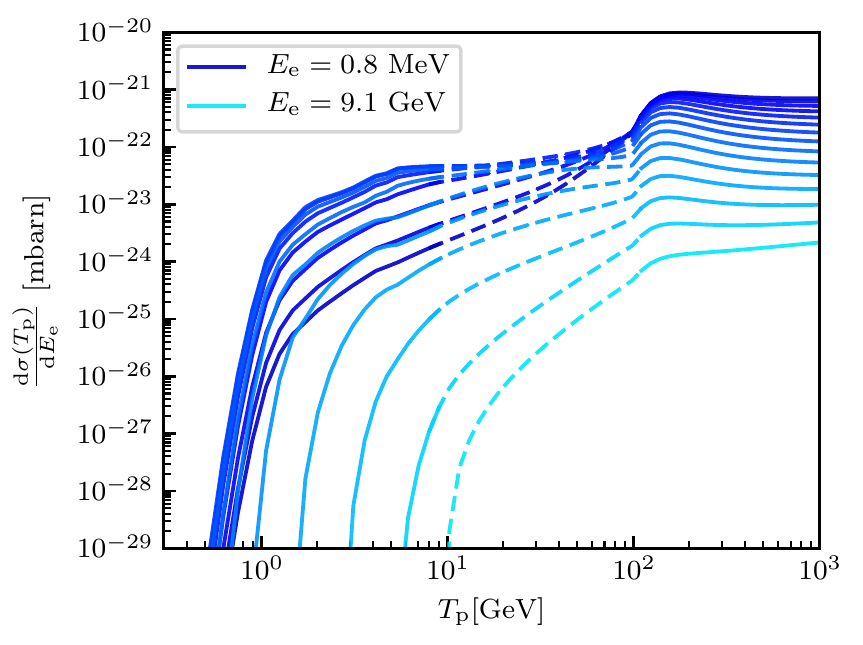}
%\par\end{centering}
\caption{Interpolation (dashed lines) of the differential cross section $\mathrm{d}\sigma_{\mathrm{e}}(E_{\mathrm{e}},E_{\mathrm{p}})/\mathrm{d}E_{\mathrm{e}}$ as a function of kinetic proton energy $T_{\mathrm{p}}$ for fixed electron energies as indicated by different colors ranging from 0.8 MeV to 9.1 GeV (in steps of equidistant momentum bins in log-space). The parametrization below $T_\rmn{p}=10$ GeV is given by \citet{2018Yang}, whereas above 100 GeV, we use \citet{2006PhRvD..74c4018K}.}
\label{fig:Interpolation-of-dsigma_e_dE}
\end{figure}

Figure~\ref{fig:Interpolation-of-dsigma_e_dE} shows the parametrizations of the differential cross sections for different electron energies. We use a cubic spline to interpolate between the \citet{2018Yang} model at low proton energies ($T_{\mathrm{p}}<10\,$GeV) and the \citet{2006PhRvD..74c4018K} model at high proton energies ($T_{\mathrm{p}}>100\,$GeV).

So far, those parametrizations only consider pp-interactions. In addition, interactions of heavier CR nuclei with the ISM have to be taken in account. Following the wounded nucleon model \citep{1976Biallas}, one can rewrite the pion production so that it can be considered to be produced in pp-collision that are enhanced by a multiplicative factor (the `nuclear enhancement factor'). Based on parametrizations by \cite{1963Lebedev} and \cite{1976OrthBuffington}, \cite{1986A&A...157..223D} finds a value of 1.39, that increases to 1.45 if heavier nuclei than helium are included, whereas \cite{1981StephensBadhwar} estimate it to 1.6 $\pm$ 0.1. Hence, we adopt the geometric mean of 1.5 and note on the relevance for studying those effects of heavier nuclei on the resulting spectrum of secondary electrons and positrons in more detail in the future \citep[using e.g., the parametrizations of sub-threshold pion production given by][]{2016Kafexhiu}.

\subsection{Parametrizations for $\sigma_{\pi}$} \label{subsec:parametrizations_for_sigma_pi}

For the total cross section of pion production, $\sigma_{\pi}$, \cite{2018Yang} use their own fit to the experimental data below 2 GeV and a different prescription given in \citet{2001PAN....64.1841G} at  higher proton energies (see red dashed lines in Fig.\ \ref{fig:Total-cross-sections}). For $T_{\mathrm{p}}>2\,\mathrm{GeV}$, they express the  cross section as 
\begin{align}
\sigma_{\pi}=\sigma_{\mathrm{p}\mathrm{p}}^{\mathrm{inel}}\left\langle n_{\pi}\right\rangle .\label{eq: sigma_pi = sigma_inel * n_pi}
\end{align}
Here, the pion average yield is parametrized as $\left\langle n_{\pi}\right\rangle =0.78(w-2)^{3/4}w^{-1/4}-1/2+\varepsilon$
where $w=\sqrt{s}/m_{\mathrm{p}}c^{2}$, $s$ denotes the square of the total energy in the CMS, i.e. $\sqrt{s}=[2m_{\mathrm{p}}c^2(E_{\mathrm{p}}+m_{\mathrm{p}}c^2)]^{1/2}$, and $\varepsilon=0$ for $\pi^{-}$,
$1/3$ for $\pi^{0}$ and $2/3$ for $\pi^{+}$. This fit is based on
experimental data by \citet{2001PAN....64.1841G}.  The resulting
curves of the total cross sections are shown in Fig.~\ref{fig:Total-cross-sections}, where the red dashed lines show the parametrizations used in \citet{2018Yang} and the points are the experimental data that they refer to. 
In the case of negative pions, one can clearly see the discontinuity
at $T_{\mathrm{p}}=2\,\mathrm{GeV}$, where the description changes from a fit to the data at $T_{\mathrm{p}}<2\,\mathrm{GeV}$ to the pion average yield by \citet{2001PAN....64.1841G} and the total inelastic cross section from \citet{2014PhRvD..90l3014K}:
\begin{align}
\sigma_{\mathrm{p}\mathrm{p}}^{\mathrm{inel}}(T_{\mathrm{p}}) & =\left[30.7-0.96\log\left(\frac{T_{\mathrm{p}}}{T_{\mathrm{p}}^{\mathrm{th}}}\right)+0.18\log^{2}\left(\frac{T_{\mathrm{p}}}{T_{\mathrm{p}}^{\mathrm{th}}}\right)\right]\nonumber \\
 & \times\left[1-\left(\frac{T_{\mathrm{p}}}{T_{\mathrm{p}}^{\mathrm{th}}}\right)^{1.9}\right]^{3}\,\mathrm{mbarn}.\label{eq:sigma_pp_inel Kafexhiu 2014}
\end{align}

The black dashed line shows the approach by \citet{1986ApJ...307...47D}, that only starts at $T_{\mathrm{p}}=0.95\,\mathrm{GeV}$ for $\sigma_{\pi^{-}}$ and an extrapolation to lower energies would lead to an underestimate of the cross section in comparison to the experimental data. Because \citet{2018Yang}
do not provide an expression for their fit to the data points below $2\,\mathrm{GeV}$, we determine our own fit to the data points up to $T_{\mathrm{p}}\leq1.1\,\mathrm{GeV}$
and use the parametrization by \citet{1986ApJ...307...47D} for higher proton energies, which leads to the orange solid line in Fig.\ \ref{fig:Total-cross-sections}.
For the cross section of positively charged pion production, we fitted the curve from Fig.\ 4 in \citet{2018Yang}, that is a sum of all channels leading to the production of positively charged pions. In this case, it connects relatively smoothly to the description of the cross section for $T_{\mathrm{p}}>2\,\mathrm{GeV}$ from Eq.\ (\ref{eq: sigma_pi = sigma_inel * n_pi}). We fit the cross sections for charged pions as follows.

The total cross section for negatively charged pions is fit by $\sigma_{\pi^-}[\mathrm{mbarn}]=\exp(f_1(x))$ with
\begin{equation}
f_1(x)= a_1  \ln(x/c_1)+ b_1 \ln ^2(x/c_1)  ,
\label{eq: sigma-pi-minus fit formula}
\end{equation}
where $x=T_{\mathrm{p}}/\mathrm{GeV}$, for $T_{\mathrm{p}}<1.1\ \mathrm{GeV}$, which corresponds to a momentum of $P_{\mathrm{p}}=1.8\ \mathrm{GeV/c}$. The parameters $a_1$, $b_1$, and $c_1$ are shown in Table~\ref{tab:fit-parameters}. Therefore, we can use the parametrization by \cite{1986ApJ...307...47D} for higher momenta, which is valid for values above  $P_{\mathrm{p}}>1.65\ \mathrm{GeV/c}$ but deviates from the data below  $T_{\mathrm{p}}<1.1\ \mathrm{GeV}$, see the black dashed line for $\sigma_{\pi^-}$ in Fig.~\ref{fig:Total-cross-sections}.
%a1, b1,c1 = 5.48679205, -10.44395614,   1.32187203 

Furthermore, we provide our parametrizations for $\sigma_{\pi^+}[\mathrm{mbarn}]=f_2(x)$ by
\begin{align}
\begin{split}
f_2(x)= [a_2 - b_2 \ln(x/c_2) + d_2 \ln^2(x/c_2)]\times (1-(x/c_2)^2)^3 +\\
                [e_2 - f_2 \ln(x/c_2) + d_2 \ln^2(x/c_2)]\times (1-(x/c_2)^{0.4})^3 ,
  \end{split}
 \label{eq: sigma-pi-plus fit formula}
\end{align}
where $x=T_{\mathrm{p}}/\mathrm{GeV}$, valid for $T_{\mathrm{p}}<1.95\ \mathrm{GeV}$. For higher energies, we apply the parametrization by \cite{2018Yang}.
%a, b, c, d, e, f =-1.5997e-01, -1.3570e-01,2.7219e-01,-2.9436e-02,-5.5311e+02, -5.3490e+02

\begin{table}
\caption{Shown are the fit parameters for equations (\ref{eq: sigma-pi-minus fit formula}) and (\ref{eq: sigma-pi-plus fit formula}).}
\label{tab:fit-parameters}
\begin{center}
\begin{tabular}{ccc} 
\hline
$a_1$& $b_1$ &$ c_1$\\ 
\hline
 $5.4868$ & $-10.4440$ &  $1.3219$ \\ 
\hline
\hline
$a_2$& $b_2$& $c_2$\\ 
\hline
$-1.5997\times 10^{-1}$ & $-1.3570\times 10^{-1}$ & $2.7219\times 10^{-1}$\\ 
\hline
\hline
$d_2$&$e_2$&$f_2$ \\
\hline
$-2.9436\times 10^{-2}$ & $-5.5311\times 10^{2}$ & $-5.3490\times 10^{2}$ \\
\hline
\end{tabular}
\end{center}
%\medskip
\end{table}

\subsection{Analytical approximation for the source function of secondary electrons \label{appendix:analytical approximation q_e}}

\begin{figure}
\begin{centering}
\includegraphics[scale=1]{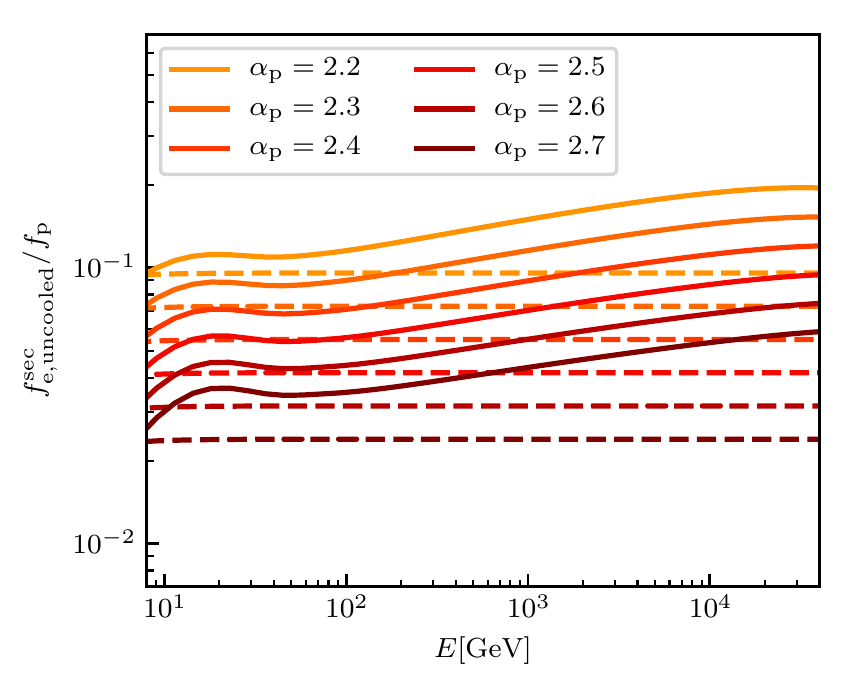}
\par\end{centering}
\caption{Ratio of secondary electrons and positrons to protons in our approach (solid lines) compared to the analytical model \citep[dashed lines,][]{2004A&A...413...17P}, for different values of the spectral index of the proton spectrum $\alpha_{\mathrm{p}}$.}
\label{fig:N_e_sec_N_p_comparison}
\end{figure}

Following \citet{2004A&A...413...17P}, here we derive an analytical approximation for the source function of secondary electrons, which provides physical insight into our numerical approach of the hadronic reaction. 
Adapting a delta approximation for the production of pions, the differential cross section of pion production reads
\begin{equation}
\frac{\mathrm{d}\sigma(E_{\pi},E_{\mathrm{p}})}{\mathrm{d}E_{\pi}}=\xi(E_{\mathrm{p}})\sigma_{\mathrm{p}\mathrm{p}}^{\pi}(E_{\mathrm{p}})\delta(E_{\pi}-\left\langle E_{\pi}\right\rangle )\theta(E_{\mathrm{p}}-E_{\mathrm{th}}).
\label{eq: pion source function simple}
\end{equation}
Assuming isospin symmetry, i.e., that the multiplicity of neutral pions is half that of charged pions, $\xi_{\pi^{0}}=(\xi_{\pi^+} + \xi_{\pi^-})/2$, this yields the pion source function
\begin{align}
q_{\pi^+}(E_{\pi}) + q_{\pi^-}(E_{\pi})&=2q_{\pi^\pm}(E_{\pi})\nonumber\\
&=\frac{2}{3} c n_{\mathrm{H}}\int\mathrm{d}E_{\mathrm{p}}f_{\mathrm{p}}(E_{\mathrm{p}})\frac{\mathrm{d}\sigma(E_{\pi},E_{\mathrm{p}})}{\mathrm{d}E_{\pi}}
\label{eq: pion source fct. simple 2}
\end{align}
from a proton energy distribution $f_{\mathrm{p}}(E_{\mathrm{p}})$.
At high energies, one can furthermore assume a constant pion multiplicity $\xi=2$, following the model by \cite{2004A&A...413...17P}, as well as a mean pion energy $\left\langle E_{\pi}\right\rangle (E_{\mathrm{p}})\simeq K_{\mathrm{p}}T_{\mathrm{p}}/\xi\simeq T_{\mathrm{p}}/(2\xi)$, where the inelasticity $K_{\mathrm{p}}$ was assumed to be $1/2$ \citep{1994A&A...286..983M}. In the high-energy limit, the proton power-law distribution in momentum is also a power-law distribution in energy since $\gamma_{\mathrm{p}}=E_{\mathrm{p}}/(m_{\mathrm{p}}c^{2})=\sqrt{1+p_{\mathrm{p}}^{2}}\approx p_{\mathrm{p}}
$ for $p_{\mathrm{p}}\gg1$ and furthermore, $T_{\mathrm{p}}/(m_{\mathrm{p}}c^{2})=\gamma_{\mathrm{p}}-1\approx\gamma_{\mathrm{p}}$.
If the energy distribution is given by a power-law with spectral index $\alpha_{\mathrm{p}}$ and normalization factor $\tilde{C}_{\mathrm{p}}$,\footnote{Note that $\tilde{C}_{\mathrm{p}}$ denotes the normalization of the CR distribution function in units of cm$^{-3}$ while $C_{\mathrm{p}}$ is the normalization of the CR source function in units of  cm$^{-3}$~s$^{-1}$.} we obtain the expression
\begin{align}
q_{\pi^+}(E_{\pi}) + q_{\pi^-}(E_{\pi}) & =\frac{4}{3}\xi^{2-a_{\mathrm{p}}}\frac{\tilde{C}_{\mathrm{p}}}{m_{\mathrm{p}}c}n_{\mathrm{H}}\sigma_{\mathrm{p}\mathrm{p}}^{\pi}(\alpha_{\mathrm{p}})\left(\frac{2E_{\pi}}{m_{\mathrm{p}}c^{2}}\right)^{-\alpha_{\mathrm{p}}}\nonumber \\
 & =\frac{16}{3}\frac{\tilde{C}_{\mathrm{p}}}{m_{\mathrm{p}}c}n_{\mathrm{H}}\sigma_{\mathrm{p}\mathrm{p}}^{\pi}(\alpha_{\mathrm{p}})\left(\frac{4E_{\pi}}{m_{\mathrm{p}}c^{2}}\right)^{-\alpha_{\mathrm{p}}}.\label{eq:pion source fct. simple 3}
\end{align}
In this approximation, the effective inelastic cross section $\sigma_{\mathrm{p}\mathrm{p}}^{\pi}$ was modeled by \citet{2004A&A...413...17P}, which also accounts for kaon decay modes. It reads
\begin{align}
\sigma_{\mathrm{p}\mathrm{p}}^{\pi}(\alpha_{\mathrm{p}})=32\times[0.96+\exp(4.4-2.4\alpha_{\mathrm{p}})]\,\mathrm{mbarn.}\label{eq: sigma_pp Pfrommer}
\end{align}
Transforming the distribution of pions into a distribution of electrons/positrons, i.e. $q_{\pi^{\pm}}\mathrm{d}E_{\pi^{\pm}}=q_{\mathrm{e}^{\pm}}\mathrm{d}E_{\mathrm{e}^{\pm}}$, and estimating the mean energy of the produced electrons or positrons from the decay channel $\pi^{\pm}\rightarrow\mathrm{e}^{\pm}+3\nu$ to be $\left\langle E_{\mathrm{e}^{\pm}}\right\rangle =\left\langle E_{\pi^{\pm}}\right\rangle /4$ \citep{1994A&A...286..983M} yields
\begin{equation}
q_{\mathrm{e}^{\pm}}^{\mathrm{sec}}(E_{\mathrm{e}^{\pm}})=q_{\pi^{\pm}}\left[E_{\pi^{\pm}}(E_{\mathrm{e}^{\pm}})\right]\frac{\mathrm{d}E_{\pi^{\pm}}}{\mathrm{d}E_{\mathrm{e}^{\pm}}}=4q_{\pi^{\pm}}(4E_{\mathrm{e}^{\pm}}).\label{eq:pion to electron distr. 2}
\end{equation}
Combining this with Eq.~(\ref{eq:pion source fct. simple 3}) gives 
\begin{align}
q_{\mathrm{e}}^{\mathrm{sec}}(E_{\mathrm{e}}) &= q_{\mathrm{e}^+}^{\mathrm{sec}}(E_{\mathrm{e}}) + q_{\mathrm{e}^-}^{\mathrm{sec}}(E_{\mathrm{e}})\nonumber\\
&=
\frac{64}{3}n_{\mathrm{H}}\frac{\tilde{C}_{\mathrm{p}}}{m_{\mathrm{p}}c}\sigma_{\mathrm{p}\mathrm{p}}^{\pi}(\alpha_{\mathrm{p}})\left(\frac{16 E_{\mathrm{e}}}{m_{\mathrm{p}}c^{2}}\right)^{-\alpha_{\mathrm{p}}}.
\label{eq:electron source fct. simple}
\end{align}
The resulting secondary electron/positron energy distribution can be inferred from the fact that the source function is a production rate that is acting on a characteristic timescale of pp-interactions $\tau_{\pi}$:
\begin{align}
    f_{\mathrm{e}^{\pm},\mathrm{uncooled}}^{\mathrm{sec}}(E_{\mathrm{e}^{\pm}})&=q_{\mathrm{e}^{\pm}}^{\mathrm{sec}}(E_{\mathrm{e}^{\pm}})\tau_{\pi},\mbox{ where}\\
    \tau_{\pi}&=\frac{1}{cn_{\mathrm{H}}K_\rmn{p}\sigma_{\mathrm{p}\mathrm{p}}^{\pi}},
    \label{eq:tau_pi}
\end{align}
provided there are no other cooling processes. Figure~\ref{fig:N_e_sec_N_p_comparison} shows the ratio of the resulting spectrum of secondary electrons plus positrons to the underlying proton spectrum as a function of energy, $E$, i.e.\ in the analytical model 
\begin{align}
    \frac{f_{\mathrm{e},\mathrm{uncooled}}^{\mathrm{sec}}(E)}{f_{\mathrm{p}}(E)} 
    = \frac{2 f_{\mathrm{e}^{\pm},\mathrm{uncooled}}^{\mathrm{sec}}(E)}{f_{\mathrm{p}}(E)}=\frac{128}{3}\, 16^{-\alpha_\mathrm{p}}.
    \label{eq:f_e,uncooled/f_p}
\end{align}

\label{lastpage}
\end{document}